\documentclass[12pt]{article}

\usepackage{amsmath,epsfig,epsf,psfrag,natbib}
\usepackage{amssymb}
\usepackage[latin1]{inputenc}
\usepackage{rotating}
\usepackage{amsbsy}
\usepackage{lscape, epstopdf}
\usepackage{color,latexsym,amsfonts,amsthm,amscd,graphicx}

\newcommand{\ben}{\begin{enumerate}}
\newcommand{\een}{\end{enumerate}}
\newcommand{\be}{\begin{equation}}
\newcommand{\ee}{\end{equation}}
\newcommand{\bas}{\begin{eqnarray*}}
\newcommand{\eas}{\end{eqnarray*}}
\newcommand{\ba}{\begin{eqnarray}}
\newcommand{\ea}{\end{eqnarray}}

\newtheorem{theorem}{Theorem}

\newtheorem{example}{Example}
\newtheorem{problem}{Problem}
\newtheorem{remark}{Remark}
\newtheorem{definition}{Definition}

\newtheorem{condition}{Condition}
\newtheorem{prop}{Proposition}

\newcommand{\e}{ { \mathbb{E}}}
\newcommand{\var}{ {\mathbb{V}\rm ar }}
\newcommand{\cov}{ {\mathbb{C}\rm ov}}

\newcommand{\pr}{{\rm pr}}

\def\T{{ \mathrm{\scriptscriptstyle \top} }}
\newcommand{\convergeto}{ {\overset{d}{\longrightarrow \; }}}

\newcommand{\zzz}{ {\rm ZZZ}}
\newcommand{\chim}{ {\rm CHIM}}

\newcommand{\ipw}{ {\rm IPW}}
\newcommand{\sipw}{ {\rm SIPW}}
\newcommand{\elw}{ {\rm ELW}}

\usepackage{authblk}
\usepackage{algorithm}
\usepackage{algorithmic}

\begin{document}
\date{}
\title{
Biased-sample empirical likelihood weighting:
an alternative to inverse probability weighting
}

\author[1]{Yukun Liu}
\author[2]{Yan Fan\thanks{Corresponding author:  fanyan212@126.com}}
\affil[1]{
    KLATASDS-MOE,
 School of Statistics,
  East China Normal University,
    Shanghai 200062, China}
\affil[2]{ School of Statistics and Information, Shanghai University of
    International Business and Economics, Shanghai 201620, China}
\renewcommand*{\Affilfont}{\small }
\renewcommand\Authands{ and }
\date{}
\maketitle

\begin{abstract}
Inverse probability weighting (IPW)  is  widely used
in many areas when data are subject to  unrepresentativeness,
missingness, or selection bias.
An inevitable challenge  with the use of IPW is that
the IPW estimator can be remarkably unstable
if some probabilities are very close to zero.
To overcome this problem,  at least three remedies
have been developed in the literature:
stabilizing,  thresholding, and trimming.
However the final estimators are still IPW type estimators,
and inevitably inherit certain weaknesses of the naive IPW estimator:
they  may still be unstable or  biased.
We propose a biased-sample empirical likelihood weighting (ELW)
method to serve the same general purpose as IPW,
while  completely overcoming  the instability of IPW-type estimators
 by circumventing the use of inverse probabilities.
The ELW weights are always well defined and  easy to implement.
We show theoretically   that the ELW estimator is asymptotically normal
and more efficient than the IPW estimator and its stabilized version
for missing data problems and unequal probability sampling without replacement.
Its asymptotic normality is also established under
 unequal probability sampling with replacement.
Our simulation results and a real data analysis
indicate that the ELW estimator is  shift-equivariant,  nearly unbiased,
and usually outperforms the IPW-type estimators in terms
of mean square error.

\vspace{1cm}

\end{abstract}

\section{Introduction}

Inverse probability weighting (IPW) has long been accepted
as the standard  estimation  procedure under unequal probability samplings
with and without replacement ever since  the work of \cite{Hansen1943} and \cite{Horvitz1952}.
 IPW  always produces an  unbiased or asymptotically unbiased estimator
with an elegant expression, regardless of the complexity
of the underlying sampling plan, and this method therefore
enjoys great popularity.
As well as survey sampling, it has  been widely used
in many other areas,   including  missing data problems
\citep{Robins1994,Wooldridge2007,Tan2010,Kim2014},
treatment effect estimation  or program evaluation
\citep{Rosenbaum1983,Rosenbaum2002,Imbens2009,Hirano2003,Cattaneo2010,Young2019,Zhao2019,Tan2020},
personalized medicine \citep{Zhang2012,Jiang2017},
and survival data analysis  \citep{Robins1992,Robins1993,
Bang2000,Ma2011,Dong2020},
where  IPW  is renamed  inverse probability of censoring weighting.
In recent years,   accompanied by optimal subsampling,
the IPW method has also  proved to be an effective approach to
validate statistical inferences for  big data
\citep{Wang2018,Wang2019,Yu2020}.

Roughly speaking,  these applications of the IPW method
can be categorized into  the following  three representative types.
\newcounter{myproblem}
\setcounter{myproblem}{\value{problem}}
\setcounter{problem}{0}
\def\theproblem{\Alph{problem}}
\begin{problem}[Missing data problem] \label{problem-a}
 Let $Z=(Y, X),$ with $Y$  being a response variable that is subject to missingness
 and $X$ an always-observed covariate.
Denote by $D$   a non-missingness indicator, with
$D=1$ if  $Y$ is observed and 0 otherwise.
Suppose $(D_i, D_iY_i, X_i),$ $i=1, 2, \ldots, N,$ are
$N$  independent and identical copies of $(Y, X, D)$.
The goal is to estimate $\theta = \e(Y)  $.
\end{problem}

\begin{problem}[Unequal probability sampling  without replacement,  UPS-WOR]\label{problem-b}
Let $\{ Z_1, \ldots, Z_N \}$   be a population of interest with known size $N$
and let $\{ (D_i, D_iZ_i): i=1, 2, \ldots, N \}$ be a sample drawn
by an UPS-WOR. The goal is to estimate the population mean  $\theta = (1/N)\sum_{i=1}^N  Z_i $.
\end{problem}
\begin{problem}[Unequal probability sampling  with replacement,  UPS-WR]\label{problem-c}
The population and the parameter
of interest are the same as in Problem \ref{problem-b},
except that the sample is drawn by  a UPS-WR.
\end{problem}
\setcounter{problem}{0}

\noindent
Special cases of Problem \ref{problem-a} include treatment effect estimation
in the framework of  Rubin's potential outcomes \citep{Rubin1974},
as well as program evaluation  in economics and other social sciences.
In a broad sense,  estimation of optimal treatment regime  and
survival data analysis both belong to  Problem \ref{problem-a}.
Poisson sampling, a popular unequal probability sampling method,
can also be regarded as  a special case of Problem \ref{problem-a}.
See Section \ref{sec3}. Optimal subsamplings in the context of  big data
are special cases of Problems \ref{problem-b} and \ref{problem-c}.

Through weighting the observations by the reciprocal of a certain
probability of inclusion in the sample,
the IPW estimator is able to account for unrepresentativeness,
missingness or selection bias
caused by non-random lack of  information
or  non-random selection of observations.
However,
the IPW estimator can be highly unstable
if there are extremely small  probabilities,
which can result in   biased estimation or
poor finite-sample performance of
the accompanying asymptotic-normality-based inference
\citep{Busso2014,Kang2007,Robins2007,Imbens2009,Cao2009,Han2019}.
As pointed out by  \cite{Robins2007}
with regard to double-robust estimators (which are IPW-type estimators)
in missing data problems,  `Whenever the ``inverse probability'' weights are
highly variable, \dots,
a small subset of the sample will have extremely large
weights relative to the remainder of the sample.
In this setting, no estimator of the marginal mean   $\mu = \e(Y)$
can be guaranteed to perform well.'
In casual inference with observational studies,
this is the well-known  limited- or non-overlap problem  in covariate distributions
in different treatment groups \citep{Crump2009,Khan2010,Yang2018}.
The IPW estimator   becomes inflated disproportionately or even breaks down
in survival analysis when the number of patients at risk
in the tails of the survival curves of censoring times is too
small \citep{Robins2000,Dong2020}.
To guarantee that the IPW estimator  possesses  consistency,
asymptotic normality, and satisfactory finite-sample performance,
it is usual to impose an unnatural lower boundedness assumption on the probabilities
\citep{Rosenbaum1983,Mccaffrey2013,Sun2018},
although tiny probabilities are frequently encountered in practice,
  especially when the propensity scores
are estimated from data
\citep{Yang2018,Ma2020}.

To overcome this notorious problem,
 at least three remedies  have been proposed in the literature:
 stabilizing,  thresholding, and trimming.
The stabilizing method
\citep{Hajek1971} rescales the IPW estimator
so that the weights sum to 1
\citep{Kang2007}.
Although straightforward,  it can often sharply reduce
the instability of the IPW estimator.
The thresholding method, proposed by
\cite{Zong2019} for solving Problem \ref{problem-b},
replaces those probabilities that are less than a given threshold by
that threshold while keeping  others  unchanged.
The parameter of interest is then estimated  by IPW with the modified probabilities.
\cite{Zong2019} proposed an easy-to-use threshold determining
procedure and showed that, in general, the resulting IPW estimator  works better
than the naive IPW estimator.
This method can  reduce the negative effect
of highly heterogeneous inclusion probabilities,
and  hence leads to improved estimation efficiency, although
at the cost of an estimation bias.

The trimming method   excludes those observations with
probabilities less than a given threshold  or, equivalently, sets their weights to zero
\citep{Crump2009,Sasaki2018}.
 However, the exact amount of trimming  is usually ad hoc and
 can affect the performance of the IPW estimator and the corresponding confidence
interval in nontrivial ways.
\cite{Ma2020} systematically investigated
the large-sample behaviour of the IPW estimator  after trimming
and found it to be sensitive to the choice of  trimming threshold
and subject to a non-negligible bias.
They  proposed a bias-corrected and trimmed IPW estimator
with an adaptively trimming threshold,
and a Studentized bootstrap procedure for  interval estimation.
Their estimator  was shown to be insensitive to small probability
weights as well as being able to correct  the bias caused by trimming.
The  bias correction technique is built on local polynomial
regressions \citep{Fan1996}, which further require a bandwidth.
Inappropriate choices of the trimmed threshold and the bandwidth
may affect the performance of their estimator.
More importantly, the bias correction technique depends on
the target quantity (e.g. $Y$ in Problem \ref{problem-a}) to be weighted,
which makes their method inapplicable to weighted optimization problems,
such as optimal treatment regime estimation \citep{Zhang2012}.

The final point estimators  of the stabilizing, trimming, and thresholding methods
are all based on IPW,
although they  adopt  different strategies
to reduce the detrimental effect of extremely small probabilities.
These IPW-type estimators inevitably inherit certain weaknesses of the naive IPW estimator:
they are either still unstable or  biased.
Also, the accompanying intervals,
regardless of whether they are asymptotic-normality-based or resampling-based,
often exhibit much undercoverage.
  See our simulation results in Section \ref{sec4}.

In this paper, we propose a biased-sample empirical likelihood weighting (ELW) estimation method
to serve the same general purpose as IPW in handling incomplete or biased data while   overcoming its instability.
We systematically investigate its finite- and large-sample properties
in the context of missing data problems  and
unequal probability samplings with and without replacement.
The proposed ELW  estimation method has several advantages over the IPW-type methods
and the usual EL method.
\ben
\item
The ELW method circumvents the use of inverse probabilities
and therefore never suffers from extremely small or
even zero selection probabilities.
It takes the maximum empirical likelihood estimates
of the probability masses of a multinomial distribution as weights,
which always range from 0 to 1.
This is the most significant advantage of the ELW method over
IPW and its variants.

\item
The ELW weights   are always well defined.
By contrast, the usual empirical likelihood weights suffer from
the  well-known convex hull constraint or the empty-set problem:
they  are undefined if the origin lies outside
the convex hull of certain transformed data points
\citep{Tsao2004,Chen2008,Liu2010}.

\item
Like the stabilized IPW estimator,
the   ELW weights always sum to 1,
which gives the ELW estimator  the nice property of  shift-equivariance.
Unfortunately,  the naive IPW estimator,
 the trimmed IPW estimator of \cite{Zong2019}, and
the IPW estimator of \cite{Ma2020}  are all sensitive to
a location shift in the response  or the parameter of interest.

\item
The ELW weights are very convenient to calculate.
Their calculation involves only solving a univariate rational
equation, which can be done efficiently  by the commonly used
bisection algorithm.
In contrast to the IPW estimator of \cite{Ma2020},
the ELW estimator is free of any tuning parameter  and is
hence more computationally efficient.
The ELW weights depend only on the propensity scores
and the total sample size, and therefore
the ELW method is directly applicable to weighted optimization problems.

\item
As we shall show,  the ELW estimator  is theoretically more efficient
than the IPW estimator in at least two scenarios:
missing data problems and unequal probability samplings without replacement.
Treatment effect estimation in observational studies
under the potential outcome framework  of \cite{Rubin1974}
can be regarded as a special case of missing data problems.
This is a bonus of ELW, since   the construction of the ELW weights
makes use of side information.
Under unequal probability sampling  with  replacement,
we cannot tell which of the ELW and IPW estimators wins theoretically.
Nevertheless, our simulation results indicate that
the ELW estimator often has
smaller mean square errors and
the accompanying interval
has better coverage accuracy in most cases.
\een

A  crucial requirement of
  ELW  is  knowledge of  the size of the finite population of interest
or a larger independent and identically distributed sample
that includes  the observed data as a subsample.
This  is also required by the
original IPW method and some of its variants,
and is  available  in most situations.
For example, in missing data problems,
the size of the overall dataset is clearly known, and
in unequal probability sampling problems,
the size of the finite population
from which the sample was drawn is usually known
a priori, since we need to construct  a sampling frame before sampling.
This mild requirement implies that the ELW method has many
potential applications beyond missing data problems, sample surveys
and  casual inference.

The remainder of this article is organized as follows.
In Section \ref{sec2},  we introduce the ELW method
by studying Problem \ref{problem-a},
namely estimating the population mean of a response
when  data are subject to missingness.
In Section \ref{sec3},  we extend the ELW method to
unequal probability sampling by
solving Problems \ref{problem-b} and   \ref{problem-c}.
A simulation study and a real-life data analysis
are conducted in Sections  \ref{sec4} and \ref{sec5}
to demonstrate the usefulness and advantage of the ELW method.
Section \ref{sec6} concludes with some discussion.
All technical proofs can be found in the supplementary material.

 \section{Empirical likelihood weighting }\label{sec2}

For ease of exposition, for the time being, we assume that the inclusion probability  or propensity score
$\pi(Z)= \pr(D=1|Z)$ is completely known and always positive,
although our method allows  $\pi(Z)$ to take zero values.
The case with unknown propensity score is considered in Section \ref{estimated-ps}.
We take the parameter of interest to be
\(
\theta = \e\{ g(Z) \}
\)
for a user-specific function $g(\cdot)$, for example $g(Z)=Y$ in Problem \ref{problem-a}.

Denote the data by $\{(D_i, D_iZ_i), i=1, 2, \ldots, N\}$,
with $Z_i = (Y_i, X_i)$
or simply $\{z_i, i=1, 2, \ldots, n\}$,
where $z_i=(y_i, x_i)$ and $n =\sum_{j=1}^N D_j$;
the covariates $X_i$ with $D_i=0$
do not come into play in most  of this paper.
The data   $\{z_i, i=1, 2, \ldots, n\}$
is in fact a biased sample of the underlying population if all $\pi(z_i)$ are not equal.
The standard IPW estimator of $\theta$ is
\ba
\label{ipw-estimator}
\hat \theta_\ipw = \frac{1}{N} \sum_{i=1}^N  \frac{ D_i }{\pi(Z_i)} g(Z_i) =
 \frac{1}{N} \sum_{i=1}^n  \frac{  g(z_i) }{\pi(z_i)}.
\ea
This expression for the IPW estimator indicates that
it becomes extremely unstable when some of the $\pi(Z_i)$ with $D_i=1$ are close to zero,
and that  the terms with  $D_i = 0$ actually contribute nothing to it.
The H\'ajek estimator,
or    stabilized IPW (SIPW)  estimator,
\bas
\hat \theta_\sipw =  \frac{ \sum_{i=1}^N    D_ig(Z_i) / \pi(Z_i) }{ \sum_{j=1}^N   D_j/\pi(Z_j)}
= \frac{ \sum_{i=1}^n    g(z_i)  / \pi(z_i) }{ \sum_{j=1}^n  1/\pi(z_j)}
\eas
replaces $N$ by $ \sum_{j=1}^N   D_j/\pi(Z_j)$, so that the weights of $g(Z_i)$ sum to 1
and the estimator becomes more stable.
Since  the size  $N$   is known,   the  zero-value $D_i$ together
with the other single-value  $D_i$
contain information about $\e(D) = \e\{ \e(D|Z) \} = \e\{\pi(Z)\}$.
The IPW estimator and its variants ignore  such side information,
and   are not able to utilize it as well,
and they consequently have potential losses of efficiency.
As a popular and flexible non-parametric technique,
 empirical likelihood  (EL) \citep{Owen1988, Owen1990,Owen2001}
can conveniently and efficiently make use of  side information to
achieve  improvements in efficiency.
This motivates us to develop a biased-sample empirical likelihood   weighting  (ELW)
estimation method to serve the same purpose as the IPW estimator,
while  overcoming its instability and improving its estimation efficiency.

\subsection{ELW estimator \label{section-elw}}

Let the distribution function  of $Z$  be $F(z)=\pr(Z\leq z)$,
where the inequality holds element-wise for vector-valued $Z$.
To estimate $\theta =  \e\{ g(Z) \} = \int g(z)\, dF(z)$,
it suffices to estimate  $F(z)$.
We consider the problem of estimating $F$ by discarding those $Z_i$ with $D_i=0$,
although these quantities may be  partially accessible.
The likelihood based on   the remaining data  is
\ba
\label{likelihood}
 \tilde L= \prod_{i=1}^N   \{   \pi(Z_i) \pr(Z_i)  \}^{D_i}(1-\alpha)^{1-D_i},
\ea
where $\alpha = \pr(D=1)   = \e\{\pi(Z)\}$.
In the principle of EL,
we model $F(z)$  by a discrete distribution function
 $F(z) = \sum_{i=1}^N p_i I(Z_i\leq z)$.
This function is not well defined, because
those $Z_i$ with $D_i=0$ are not fully available.
Nevertheless, our final estimator, say $\hat p_i$, of $p_i$
satisfies $\hat p_i=\hat p_i D_i$, making the resulting maximum EL
estimators of $F$ and $\theta$
 well-defined statistics.

With $p_i$ in place of $\pr(Z_i)$ in \eqref{likelihood}
and taking logarithms, we have  the empirical log-likelihood
\ba
\label{log-el}
\tilde \ell =  \sum_{i=1}^N  [ D_i \log(p_i)+ D_i \log\{  \pi(Z_i) \} + (1-D_i)\log (1-\alpha ) ].
\ea
Those $p_i$ that are feasible satisfy
\ba
\label{constr-pi}
p_i\geq 0, \quad \sum_{i=1}^N  p_i=1, \quad
\sum_{i=1}^N p_i \{ \pi(Z_i) - \alpha \} =0.
\ea
We  emphasize that  although  those $Z_i$ with $D_i=0$
appear in $\tilde \ell$  and $\sum_{i=1}^N  p_i \{ \pi(Z_i) - \alpha \} =0$,
they  have no likelihood contribution or
any influence on the resulting EL method.

The proposed  EL estimator   of $F(z)$,
or equivalently of the $p_i$, is obtained by maximizing
the empirical log-likelihood \eqref{log-el}
subject to \eqref{constr-pi}.
For fixed $\alpha$,
the maximum of the log-EL in \eqref{log-el} subject to \eqref{constr-pi}
is attained at
\ba
\label{pi-expression}
p_i = \frac{1}{n }  \frac{D_i}{1+\lambda(\alpha) \{ \pi(Z_i) - \alpha \} },
\ea
where  $\lambda(\alpha)$ satisfies
\ba
\label{eq-lambda}
\frac{1}{n } \sum_{i=1}^N  \frac{D_i}{1+\lambda(\alpha) \{ \pi(Z_i) - \alpha \} } \{ \pi(Z_i) - \alpha \} = 0.
\ea
Putting \eqref{pi-expression} into
\eqref{log-el} gives  the profile log-EL of $\alpha$  (up to a constant that is independent of $\alpha$)
\bas
 \ell(\alpha)
 =
 \sum_{i=1}^N  \{ - D_i \log[ 1+\lambda(\alpha)\{ \pi(Z_i) - \alpha \}]  +   (1-D_i)\log (1-\alpha ) \}.
\eas
This immediately gives  $\hat \alpha = \arg\max \ell(\alpha)$,
the EL estimator of $\alpha$.
Accordingly, the EL estimators of $p_i$  and $F(z)$ are
\ba
\label{hat-p}
\hat p_i  = \frac{1}{ n}  \frac{D_i}{1+ \lambda (\hat \alpha) \{ \pi(Z_i) - \hat \alpha \} }
\ea
and  $\hat F (z) = \sum_{i=1}^N  \hat p_i I(Z_i\leq z)$.
Finally, the EL estimator or  the  ELW estimator of $\theta$
is
\ba
\label{gee-w}
\hat \theta_{\elw} =\int g(z) d\hat  F(z) = \sum_{i=1}^N \hat p_i  g(Z_i).
\ea
Obviously, both $\hat F (z) $ and $\hat \theta$ are  well-defined statistics
 because $\hat p_i  = D_i \hat p_i$.

\begin{remark}
The focus of this paper is the derivation of a better weighting method
than IPW. We achieve this by taking the  probability masses of
the maximum empirical likelihood estimator of $F(z)$
as weights.
Our ELW method requires  the total sample size $N$ and
the  (estimated) selection probabilities.
After multiplication by ${N\choose n},$  the likelihood in \eqref{likelihood}
is a degenerate case (with only one capture occasion)
of the full likelihood for capture--recapture data of \cite{Liu2017},
in which $N$ is unknown and the focus of estimation.
The case with known $N$ has been studied by \cite{Li1998}
for biased and truncated data, although their interest was not in better weighting,
but  rather in estimating the distributions of the target and truncation variables, together with
the unknown parameters involved.

\end{remark}

\begin{remark}
When the propensity scores are available for those with $D_i=1,$
the ELW method works even if  the responses $Y_i$ and
the covariates $X_i$ are missing simultaneously
for all data $Z_i=(Y_i, X_i)$ with $D_i=0$.  However,
it may be subject to loss of efficiency  when  all covariates are observed.
To improve efficiency in this case,  we may incorporate estimating equations such as
$\sum_{i=1}^N  p_i\{h(X_i) -  \bar h\} = 0$ in the definition of the empirical log-likelihood
\eqref{log-el}, where $h$ is a user-specific function and $\bar h = (1/N)\sum_{i=1}^N h(X_i)$.
This strategy was also adopted by  \cite{Qin2008} to improve efficiency
and by \cite{Han2013} and \cite{Han2014} to construct  multiple robust  estimators.
The price paid for this modification, however, is that  the resulting ELW weights  are undefined
if  $\bar h$ lies outside  the convex hull of $\{h(X_i): D_i=1, i=1, 2,\ldots, N\}$
\citep{Tsao2004,Chen2008,Liu2010}.
The probability that the ELW weights
are undefined can be large when the sample size is small
and/or  the dimension of $h$ is high \citep{Tsao2004};
the calculational  burden of the ELW weights will also be  heavier.
As the focus of this paper is on the development of an estimation method that is
robust to small probabilities, we do not incorporate side information
in the proposed ELW method for the time being.
\end{remark}

\subsection{Practical implementation}

The key to calculating  the proposed EL estimators, including
the EL estimator $\hat F$ of $F$ and the ELW estimator $\hat \theta_{\elw}$,
is to calculate $\hat \alpha$ by maximizing $\ell(\alpha)$.
This necessitates a double iterative algorithm  because
 $\ell(\alpha)$  involves an implicit function $\lambda(\alpha)$,
 and thus it seems to be rather a difficult task.
We find a more convenient solution, in which we need  only
solve a univariate equation.

Mathematically,  $\hat \alpha = \arg\max \ell(\alpha)$ is  a solution to
\ba
\label{eq-alpha}
 0 &=&
    \sum_{i=1}^N  \left[ \frac{  D_i   \lambda }{
    1+\lambda\{ \pi(Z_i) - \alpha \} }  -    \frac{1-D_i} {1-\alpha } \right].
\ea
Combining  \eqref{eq-lambda} and \eqref{eq-alpha} gives
\ba
\label{lambda-alpha}
\lambda = \frac{ N - n }{n (1-\alpha)}.
\ea
Putting this expression into \eqref{eq-lambda} leads to an equivalent equation
for $\alpha$:
\ba
\label{eq-alpha-equivalent2}
    \sum_{i=1}^N   \frac{D_i(\pi(Z_i) - \alpha) }{
  n /N +(1-n /N)\pi(Z_i)  - \alpha   }
  &=& 0.
\ea
As  \eqref{eq-alpha-equivalent2}   has multiple roots,
it is necessary to identify the interval containing the desired root.
Denote   the observed $Z_i$  by $z_1, \ldots, z_{n}$
and define \(
\xi_i =  n /N +(1-n /N) \pi(z_i)
\) for $i=1, 2, \ldots, n$.
Equation \eqref{eq-alpha-equivalent2} is further equivalent to
$K(\alpha) = 0$, where
\(
K(\alpha)
 =   \sum_{i=1}^{n }   \{\pi(z_i) - \alpha \}/(\xi_i  - \alpha )
\).
Because  $\alpha \in (0, 1)$, $\xi_i\geq n /N$,
 and $n /N$ is  a consistent estimator of $\alpha$,
the desired root of $K(\alpha) = 0$
 should lie between $0$ and $\min \xi_i$.
Actually, there must exist one and only one solution
to $K(\alpha) = 0$ between $0$ and $\min \xi_i$.
Because $\xi_i\geq \pi(z_i)$,
it follows that   $K\{ \min \pi(z_i)\}\geq 0$,
$\lim_{\alpha \uparrow  \min \xi_i} K(\alpha) = -\infty$,
and that $K(\alpha) $ is  strictly decreasing between $0$ and $\min \xi_i$.
By the intermediate value theorem, there must exist one and only one solution,
denoted by $\hat \alpha$,  in  $[\min \pi(z_i), \min \xi_i)$ such that $K(\hat \alpha) = 0$.
It is worth noting that if all the $\pi(z_i)$ are equal
and equal to $\alpha_0$, then
$\hat \alpha = \alpha_0$  and the resulting $\hat p_i$ are all equal to $1/n$,
and the ELW   estimator reduces to the sample mean,
i.e., $\hat \theta_{\elw}   = (1/n) \sum_{i=1}^n g(z_i)$.
Otherwise,   all $\pi(z_i)$ ($i=1, 2, \ldots, n$) are not equal to each other,
and $\hat \alpha$, $\hat p_i$, and  $\hat \theta_{\elw}$ are all non-degenerate.

The proposed ELW estimation procedure can be implemented by Algorithm \ref{elw-procedure}.
The   second step can be efficiently achieved by a bi-section search algorithm,
and  the remaining steps all involve only closed-form calculations,
making   the ELW procedure extremely   easy to implement.
\begin{algorithm}[h]
\caption{ELW estimation procedure \label{elw-procedure}}

\renewcommand{\algorithmicrequire}{\textbf{Input:}} 
\renewcommand{\algorithmicensure}{\textbf{Output:}} 
\begin{algorithmic}

\REQUIRE
The missing dataset  $\{ (D_i, D_i g(Z_i), \pi(Z_i)): i=1, 2, \ldots, N \}$.

\ENSURE
The ELW estimate, $\hat \theta_{\elw}$, of $\theta = \e\{g(Z)\}$.

\STATE   {\bf Step 1.}
Calculate  $n=\sum_{i=1}^ND_i$,
 $\zeta_l = \min\{\pi(Z_i): D_i=1, i=1, 2, \ldots, N \}$
 and $\zeta_u = n/N + (1-n/N) \zeta_l$.

\STATE   {\bf Step 2.}
Calculate $\hat \alpha$   by solving    \eqref{eq-alpha-equivalent2}
  in the interval $[\zeta_l, \zeta_u)$,
  and calculate $  \lambda(\hat \alpha)   =  ( N - n )/\{ n (1- \hat \alpha) \}$.

\STATE  {\bf Step 3.}
Calculate
$\hat p_i  = D_i n^{-1}  [ 1+ \lambda (\hat \alpha) \{ \pi(Z_i) - \hat \alpha \} ]^{-1}$
for $i=1, 2, \ldots, N$.

\STATE  {\bf Step 4.}
Calculate $ \hat \theta_{\elw} = \sum_{i=1}^N \hat p_i D_i g(Z_i)$.

\end{algorithmic}
\end{algorithm}

\subsection{Finite- and  large-sample properties}

The non-zero  EL weights are
\(
(1-\hat \alpha)/\{N(\xi_i-\hat \alpha)\}
\) for $ 1\leq i\leq n$.
We use the maximum  weight ratio
\(
\kappa  =
 (\max_{1\leq i \leq n } \xi_i-\hat \alpha)/(\min_{1\leq i \leq n } \xi_i-\hat \alpha)
\)
among the non-zero EL weights
to quantify the dispersion between the EL weights.
The following
proposition establishes an upper bound  on  $\kappa $.

\begin{prop}
\label{prop-kappa}
Suppose $\pi(z_i)$ $(1\leq i\leq n )$ take
$m$ $\geq 2$ distinct values $\pi_{(1)} < \ldots<\pi_{(m)}$ $(m\geq 2)$.
  If there exists  $\varepsilon \in (0, 1)$ such that  $\pi_{(m)}-\pi_{(1)}>\varepsilon$
and $n_1/n<1-\varepsilon,$
then
 $\kappa  \leq n  /\varepsilon^3 $.
\end{prop}

Proposition \ref{prop-kappa} indicates that the ELW method works
even if the smallest $\pi(z_i)$ is as small as zero.
However, the maximum weight ratio of the IPW estimator,
namely $\max_{1\leq i\leq n }\pi(z_i)/\min_{1\leq i\leq n }\pi(z_i)$,
has no such a guarantee,
and the IPW estimator becomes extremely unstable
when some of the $\pi(z_i)$ are close to zero.
In particular,  it fails to work when   $\min_{1\leq i\leq n}\pi(z_i) $ is exactly zero.
Our ELW estimator successfully and completely overcomes this issue,
which is its most significant advantage over
the traditional IPW  in   finite-sample performance.

Next, we show that asymptotically our ELW estimator is
unbiased and  more efficient than the IPW estimator.
This is a bonus of using ELW,  and also a significant advantage that it has  over
the conventional  IPW  in  large-sample performance.
For ease of presentation,   let $B_{11} =   \e\{  1  /\pi(Z )  \}$,
and, for two generic functions $f$ and $h$,   define
$
B_{fh} =    \e\{  f(Z) h^\T (Z) /\pi(Z )  \}
$,
$B_{f1} =   \e\{  f(Z)  /\pi(Z )  \}$,
and $ G  =   (B_{g1} -     \theta_0 )/(  B_{11} - 1 )$.
We   denote $A^{\otimes 2} = AA^\T$ for a vector or matrix $A$.

\begin{theorem}
\label{CLT-A-known}
Let $  \alpha_0 \in (0, 1)$ be the truth of $\alpha$.
Suppose that $\var\{ \pi(Z)|D=1 \}>0$ and
that $ B_{11} $ and $ B_{gg} $ are both finite.
As $N$ goes to infinity,
\ben
\item[(a)]
$
\sqrt{N}(\hat \theta_{\elw} - \theta_0)
\convergeto N(0, \Sigma_{\elw}),
$
where
\bas
\Sigma_{\elw}
&=&
\var
\left\{
\left( \frac{D   g(Z)}{   \pi(Z  )   }- \theta_0\right)
-
G     \frac{  D   -\pi(Z )   }{  \pi(Z  )     }
\right\}
=
 ( B_{gg} - \theta_0^{\otimes 2}) -
  G^{\otimes 2} (      B_{11} -  1);
\eas
 \item[(b)]
$
\sqrt{N}(\hat \theta_{\ipw} - \theta_0)
\convergeto N(0, \Sigma_{\ipw}),
$
where
\bas
\Sigma_{\ipw }  =
\var \left\{ \frac{D   g(Z)}{   \pi(Z, \beta_0 )   }- \theta_0\right\}
=  B_{gg} - \theta_0^{\otimes 2};
 \eas
 \item[(c)]
$
\sqrt{N}(\hat \theta_{\sipw} - \theta_0)
\convergeto N(0, \Sigma_{\sipw}),
$
where
\bas
\Sigma_{\sipw }  &=&
\var \left\{  \left(\frac{D   g(Z)}{   \pi(Z, \beta_0 )   }- \theta_0\right)
- \theta_0 \left(\frac{D}{\pi(Z, \theta_0)}  -1 \right)\right\} \\
&=&   (B_{gg} - \theta_0^{\otimes 2}) - \theta_0 (B_{g1} - \theta_0)^\T
- (B_{g1} - \theta_0)\theta_0^\T + \theta_0^{\otimes 2}(B_{11}-1);
 \eas
\item[(d)]
the ELW estimator $\hat \theta_{\elw}$
is more efficient than the IPW estimator $\hat \theta_{\ipw}$
and the SIPW estimator $\hat \theta_{\sipw},$
i.e.
\(
  \Sigma_{\elw} \leq \Sigma_{\ipw}
\)
and
\(
  \Sigma_{\elw} \leq \Sigma_{\sipw}
,\)
where the equalities hold only if $\pi(Z)$ is degenerate.
\een

\end{theorem}

The assumption $\var\{ \pi(Z)|D=1 \}>0$ guarantees that
with probability tending to 1,
the propensity scores are not all equal to each other,
and so  the ELW estimator is non-degenerate.
A reasonable estimator of $ \Sigma_{\elw}$ is required
in the construction of  Wald-type confidence intervals for $\theta$.
Inspired by the fact that $\hat p_i \approx  D_i/\{ N \pi(Z_i)\}$,
we propose to estimate $ \Sigma_{\elw}$ with the ELW method by
\ba
\label{Sigma-ELW-est}
 \widehat{\Sigma}_{\elw} =  (\hat  B_{gg} -\hat  \theta_{\elw}^{\otimes 2}) -
  (\hat B_{g1}     -      \hat  \theta_{\elw})^{\otimes 2}  /(      \hat B_{11} -  1),
\ea
 where
$
\hat B_{11} =   N  \sum_{i=1}^N (\hat p_i )^2
$,
$
\hat B_{g1} = N \sum_{i=1}^N g(Z_i)(\hat p_i )^2
$,
and
$
\hat B_{gg} =  N \sum_{i=1}^N \{g(Z_i)\}^{\otimes 2}(\hat p_i )^2.
$
It is worth stressing that the ELW-based variance estimator
is again insensitive to small probabilities, since it circumvents the use of inverse probabilities.

\subsection{Estimated propensity score}
\label{estimated-ps}

In many situations, such as missing data problems and causal inference,
the propensity score is unknown and needs be estimated from the observed data.
The ELW and IPW estimators have different large-sample behaviours
if we take  the variability of the estimated propensity score  into account.
Suppose that $\pi(\cdot)$ is parametrically modelled by  $\pi(Z, \beta)$
with true parameter value $\beta_0$.
\begin{condition}
\label{condition1-beta-un}
The following conditions are satisfied:
\ben
\item[(a)]
  The function $\pi(Z, \beta)$ is continuous in $\beta$
  and    $\pi(Z, \beta_0) = \pi(Z)$  for all $Z$.
\item[(b)]
 There exists a constant $\varepsilon>0$ such that
(1) $ \e\{  \sup_{\beta:  |\beta-\beta_0|<\varepsilon} \|\nabla_{\beta} \pi(Z, \beta )\|  \} < \infty$
and
(2)
 $  \e \{ \sup_{\beta:  |\beta-\beta_0|<\varepsilon} 1/\pi(Z, \beta) \} <\infty$.
\item[(c)]
The truth $\alpha_0$ of $\alpha$  satisfies   $ 0 < \alpha_0 < 1 $ and
$\var\{ \pi(Z, \beta_0)|D=1 \}>0$.
\een
\end{condition}
\noindent
Condition \ref{condition1-beta-un}(a) holds when
the non-missingness indicator $D$ follows  logistic and probit models,
which are commonly used in the literature.
Condition \ref{condition1-beta-un}(b) together with the other conditions guarantee
the consistency of $\hat \alpha$ and hence the consistency of the ELW estimator.
Under Condition  \ref{condition1-beta-un}(c),
with probability tending to 1,
the observed propensity scores are not all equal to each other, and
so  the ELW estimator is non-degenerate.

Under  Condition \ref{condition1-beta-un}(a),     $\pi(Z, \beta_0) = \pi(Z)$ and
therefore
$
B_{11}   = \e\{ 1/\pi(Z, \beta_0)  \}
$,
$
B_{g1} = \e\{ g(Z)/\pi(Z, \beta_0)  \}
$,  and
$
B_{gg}= \e[ \{g(Z)\}^{\otimes 2}/\pi(Z, \beta_0)   ].
$
We still use $ G  =   (B_{g1} -     \theta_0 )/(  B_{11} - 1 )$.
In addition, we denote
$
B_{ \dot \pi 1}
=   \e\{  \nabla_{\beta} \pi(Z, \beta_0)  /\pi(Z, \beta_0)  \}
$
and
$
B_{g \dot \pi} =    \e[ g(Z)  \{\nabla_{\beta} \pi(Z, \beta_0)\}^\T  /\pi(Z, \beta_0)  ].
$

\begin{theorem}
\label{CLT-A-unknown}
Assume Condition  \ref{condition1-beta-un} and that
   $\hat \beta$ satisfies
$
 \hat \beta - \beta_0
 =  N^{-1} \sum_{i=1}^N    h(D_i, Z_i)  + o_p(N^{-1/2}),
$
where the influence function $h(D, Z)$ has  zero mean.
Suppose  that $ B_{11},  B_{gg}, B_{hh},$ and $B_{\dot{\pi}\dot{\pi}}$ are all finite.
 As $N$ goes to infinity,
\ben
\item[(a)]
$
\sqrt{N}(\hat \theta_{\elw} - \theta_0)
\convergeto N(0, \Sigma_{\elw, e} ),
$
where
\bas
\Sigma_{\elw, e} =
\var
\left\{
\left( \frac{D   g(Z)}{   \pi(Z, \beta_0 )   }- \theta_0\right)
-
G     \frac{  D   -\pi(Z,  \beta_0 )   }{  \pi(Z,  \beta_0 )     }
+   (G  B_{1 \dot{\pi} }    -   B_{g \dot{\pi}})  h(D, Z)
\right\};
\eas
\item[(b)]
$
\sqrt{N} (\hat\theta_{\ipw} - \theta_0)
 \convergeto   N(0, \Sigma_{\ipw, e} ),
$ where
\bas
\Sigma_{\ipw, e}  =
\var \left\{ \left( \frac{D   g(Z)}{   \pi(Z, \beta_0 )   }- \theta_0\right)
 - B_{g\dot \pi}h(D, Z )  \right\};
 \eas
\item[(c)]
$
\sqrt{N} (\hat\theta_{\sipw} - \theta_0)
 \convergeto   N(0, \Sigma_{\sipw, e} ),
$ where
\bas
\Sigma_{\sipw, e}  =
 \var
 \left\{
\frac{D_i  g(Z_i) - D_i\theta_0  }{ \pi(Z_i,\beta_0)}
+
(\theta_0   B_{1\dot \pi} -   B_{g\dot \pi})  h(D_i, Z_i)
\right\}.
 \eas
\een

\end{theorem}

According to Theorem \ref{CLT-A-unknown},
  the ELW,  IPW and SIPW estimators still follow asymptotic normal distributions
when the propensity score involves a finite-dimensional unknown parameter.
However, in general, the efficiency gain of the ELW estimator
over the IPW and SIPW estimators is no longer guaranteed.

The most popular choice for $ \hat \beta$ is  the maximum likelihood estimator.
For example, if data are missing at random \citep{Rubin1976} and
the covariates $X_i$ are all observed,
  the maximum likelihood estimator $\hat \beta$ is the maximizer of
$\sum_{i=1}^N [ D_i \log \pi(Z_i, \beta) + (1-D_i) \pi\{1-\pi(Z_i, \beta)\} ]$.
In this case,
\bas
h(D, Z)
&=& \frac{D-\pi(Z, \beta_0 ) }{\pi(Z, \beta_0 ) \{1-\pi(Z, \beta_0 ) \}}
(\tilde B_{\dot{\pi}\dot{\pi}})^{-1}  \nabla_{\beta}\pi(Z, \beta_0 ),
\eas
where
$
\tilde B_{\dot{\pi}\dot{\pi}}
=
\e  [ \{ \nabla_{\beta}\pi(Z, \beta_0 ) \}^{\otimes 2}/\{ \pi(Z, \beta_0 ) (1-\pi(Z, \beta_0 ) )\} ].
$
The asymptotic variance of the ELW  estimator  is
\bas
\Sigma_{\elw, e}
&=&
B_{gg} - \theta_0^{\otimes 2}
-
G (B_{11}-1)G^\T
-
  (G  B_{1 \dot{\pi} }    -   B_{g \dot{\pi}}) (\tilde B_{\dot{\pi}\dot{\pi}})^{-1}
   (G  B_{1 \dot{\pi} }    -   B_{g \dot{\pi}})^\T.
\eas
Again, an ELW estimator can be constructed for
$ \Sigma_{\elw, e}$.

\subsection{Resampling-based interval estimation}\label{sec2.5}

Based on Theorems \ref{CLT-A-known} and \ref{CLT-A-unknown},
we can construct Wald-type confidence intervals for $\theta$
once a consistent estimator for the asymptotic variance is available.
The asymptotic normality of the ELW, IPW, and SIPW estimators requires
$B_{11} = \e[\{  \pi(Z_i) \}^{-1} ] < \infty$  and $B_{gg}<\infty$.
 If this is violated, the Wald-type confidence intervals may not
 have the promised coverage probability.
This  dilemma  can be overcome by resampling.
We propose to construct confidence intervals
for $\theta$ by the resampling method in Algorithm \ref{subsample}.

\begin{algorithm}[h]
\caption{Wald confidence region based on resampling and ELW \label{subsample}}

\renewcommand{\algorithmicrequire}{\textbf{Input:}} 
\renewcommand{\algorithmicensure}{\textbf{Output:}} 

\begin{algorithmic}
\REQUIRE
The missing dataset  $\{ (D_i, D_i g(Z_i), \pi(Z_i)): i=1, 2, \ldots, N \}$.
Calculate the ELW estimator $ \hat  \theta_{\elw}$ and
the proposed variance estimator $ \widehat{\Sigma}_{\elw}$, and
define
$T_N =  \sqrt{N} (\widehat{\Sigma}_{\elw})^{-1/2} (\hat  \theta_{\elw} - \theta_0)$.

\ENSURE
Wald confidence region for   $\theta = \e\{g(Z)\}$ based on resampling and ELW

\STATE   {\bf Step 1.}
Draw $M \ll N$ (e.g. $M=\sqrt{N}$) observations, say
$ (D_i^*, D_i^* g(Z_i^*), \pi(Z_i^*))$ ($ 1\leq i\leq M$),
 from the original sample
by simple random sampling without replacement.

\STATE  {\bf Step 2.}
Calculate the counterparts of $\hat \theta_{\ipw}$ and $\widehat{\Sigma}_{\elw}$
based on the subsample,
denoted by $ \hat  \theta_{\elw}^*$ and  $ \widehat{\Sigma}_{\elw}^*$ .
Construct
$T_M^* =  \sqrt{M}
 (\widehat{\Sigma}_{\elw}^*)^{-1/2} (\hat  \theta_{\elw}^* - \hat  \theta_{\elw})$.

\STATE  {\bf Step 3.}
Repeat Steps 1 and 2 $B=1000$ times and denote the resulting
test statistics by $\{ T_{M, i}^*: i=1, 2, \ldots, B\}$.
Let $t_i^* = \| T_{M, i}^*- \bar T^* \| $, where $\bar T^* = (1/B) \sum_{i=1}^B  T_{M, i}^*$.
Denote the $(1-\alpha)$ empirical quantile of the $t_i^*$  by $q_{1-\alpha}^*$.
Then a $(1-\alpha)$-level confidence
region for $\theta$ can be constructed as
\(
\{\theta:
\| \sqrt{N} (\widehat{\Sigma}_{\elw})^{-1/2} (\hat  \theta_{\elw} - \theta ) - \bar T^*  \|
\leq q_{1-\alpha}^*
 \}.
\)
\end{algorithmic}
\end{algorithm}

In the case of the estimated propensity score  $\widehat\pi(Z_i )$,
we replace  $\pi(Z_i)$ and $ \widehat{\Sigma}_{\elw}$ by
 $\widehat\pi(Z_i )$ and $ \widehat{\Sigma}_{\elw, e}$, respectively.
The ELW variance estimator  $ \widehat{\Sigma}_{\elw}$
converges in probability to $  {\Sigma}_{\elw}$,
which is assumed to be positive definite.
This, together with Theorems \ref{CLT-A-known} and \ref{CLT-A-unknown}, implies that
$T_N$ converges in distribution to the standard normal,
an obviously continuous distribution.
By Corollary 2.1 of \cite{Politis1994},
the empirical distribution of $T_M^* $ is a uniformly consistent estimator
of the distribution of $T_N$,
which is formally summarized in  Theorem \ref{validation-of-subsample}.
This validates  the interval estimator produced by Algorithm \ref{subsample}.
\begin{theorem}
\label{validation-of-subsample}
Assume the conditions in Theorem \ref{CLT-A-known}   (for a known propensity score)
or those in Theorem  \ref{CLT-A-unknown} (for an estimated propensity score)
are satisfied.
As $N\rightarrow \infty,$ if $M \rightarrow \infty$ and $M/N \rightarrow 0,$ then
$
\sup_{t\geq 0} | P(T_N \leq  t ) - P^*(T_M^* \leq t) | = o_p(1),
$
where $P^*$ is the conditional probability given
the original sample.
\end{theorem}

\section{Extension to unequal probability samplings}\label{sec3}

We have shown that the ELW method works
for missing data problems, where data are independent
and identically distributed.
The results in Theorem \ref{CLT-A-known} also hold for Poisson sampling, a special case of UPS-WOR.
Under Poisson sampling, the indicators $D_i$  are all independent
of each other and $n = \sum_{i=1}^N D_i$ is random.
Suppose there exists a function $\pi(\cdot)$  such that
$ P(D_i=1) = \pi(Z_i)$; then $\{(D_i, D_iZ_i), i=1, 2, \ldots, N\}$
can be regarded as missing data and the ELW method applies.
However, Theorem \ref{CLT-A-known} no longer holds under  more general UPS-WORs,
where the sample size is fixed and the data are correlated,
or UPS-WRs, where a piece of data may be sampled multiple times.
In this section, we extend the ELW method to UPS-WOR and UPS-WR by
solving Problems \ref{problem-b}  and \ref{problem-c}.

\subsection{ELW for UPS-WOR}

With the notation of Problem \ref{problem-b},
denote the inclusion probability of $Z_i$  by $\pi_i$.
The population size $N$  and all the $\pi_i$ are known a priori.
The IPW estimator of $\theta$ is the famous Horvitz--Thompson estimator
\citep{Horvitz1952}
\ba
\label{eq-ipw}
\hat \theta_\ipw = \frac{1}{N} \sum_{i=1}^N  \frac{ D_i}{\pi_i}  g(Z_i).
\ea
With $\pi_i$ in place of $\pi(Z_i)$,
we propose to directly apply the ELW  method from Section \ref{section-elw} and estimate
$\theta_0 = (1/N)\sum_{i=1}^N  g(Z_i)$ by  $\hat \theta_{\elw}$
in \eqref{gee-w}.  Although the ELW estimators for UPS-WOR and for missing data problems
have the same form,  their random behaviours are totally differentm because
the joint distributions of the observed data are different.

The asymptotic normality  of the IPW estimator,
although difficult,
 has been established
for many commonly used samplings,
including simple random sampling with or without
replacement \citep{Erdos1959,Hajek1960},
rejective sampling with unequal probabilities (H$\rm\acute{a}$jek, 1964),
stratified unequal probability sampling with or without
replacement  \citep{Krewski1981,Bickel1984},
and two-phase sampling \citep{Chen2007}.
See \cite{Wu2020} for a comprehensive review.

We establish the asymptotic normality of the ELW estimator
following the results of \cite{Patterson2001} and \cite{Branden2012}.
A crucial requirement  in these papers
for the series of random elements
is  linear negative dependence
or negative association,
which are defined formally as follows.

\begin{definition}[\cite{Patterson2001}]
A sequence of random variables, $\{ X_i \},$
is said to be linearly negatively dependent (LIND)  if
for any disjoint subsets of indices $A, B$
and positive constants $\lambda_j,$
\bas
P\left(\sum_{k\in A} \lambda_k X_k \leq s,  \sum_{j\in B} \lambda_j X_j \leq t  \right)
\leq
P\left(\sum_{k\in A} \lambda_k X_k \leq s  \right)\cdot
P\left( \sum_{j\in B} \lambda_j X_j  \leq t\right)
\eas
for any real numbers $s$ and $t$.

\end{definition}

\begin{definition}[\cite{Branden2012}]
A family $X=(X_1, \ldots, X_N)$ of random variables is said
to be negatively associated (NA) if
\(
\e\{f(X) g(X)\} \leq  \e\{f(X) \} \e\{  g(X)\}
\)
for all increasing functions $f$ and $g$  such that
there exists  $ I \subset \{1, 2, \ldots, N\}$
such that  $f$ depends only on  $\{X_i: i\in I\}$
and $g$ depends only on  $\{X_j: i \neq I\}$.
\end{definition}

\cite{Patterson2001} proved that
if a series of random variables is NA, then it must be LIND, and that
for LIND random series,
 asymptotic   normality of the IPW estimator holds under mild additional conditions.
\cite{Branden2012}
showed that many commonly used $\pi ps$ samplings, including conditional Poisson sampling,
Sampford sampling,  Pareto sampling, and pivotal sampling
satisfy the NA  property.
In summary, these results imply that,
under mild additional conditions,
the IPW estimator for these  commonly used $\pi ps$ samplings without replacement
follows an asymptotic normal distribution.

Suppose that there is a sequence of large datasets,
$\{Z_{\nu 1}, \ldots, Z_{\nu N_\nu}\}$,
of size $N_\nu$, $\nu=1, 2, \ldots, $
  all of which are regarded as finite sets of non-random numbers.
The large dataset under study is one of them.
For a particular large dataset  $\{Z_{\nu 1}, \ldots, Z_{\nu N_\nu}\}$,
a sample of size $n_\nu$ is taken from it with a prespecified UPS-WOR.
Define $D_{\nu k} = 1$ if $Z_{\nu k}$ is selected, and 0 otherwise for $1\leq k\leq N_{\nu}$.
Then, $n_\nu = \sum_{k=1}^{N_\nu}D_{\nu k}$.
Suppose that $N_\nu \rightarrow \infty$ and
$\alpha_{\nu0} \equiv n_\nu/N_{\nu} \rightarrow \alpha_0 \in (0, 1)$,
as $\nu\rightarrow\infty$.

\begin{theorem}
\label{CLT-B-ups}
Let  $\{Z_{\nu k}, 1\leq k\leq N_\nu, \nu\geq 1\}$ be
an array of  $q$-dimensional constant vectors
and $\{D_{\nu k}, 1\leq k\leq N_\nu, \nu\geq 1\}$
be an array of  row-wise LIND random variables.
Let
$\pi_{\nu k} = \e (D_{\nu k})$  and
$
W_{\nu k} = (g^\T(Z_{\nu k}), 1)^\T /\pi_{\nu k},$
and define
\(
S_\nu^2 =    N_\nu^{-1}\var\{ \sum_{k=1}^{N_\nu}   W_{\nu k}  ( D_{\nu k} - \pi_{\nu k})  \}.
\)
Suppose that
\bas
&(1)&
\lim_{\nu\rightarrow \infty}   S_\nu^2 =  V_0  \quad
 \mbox{ is a positive definite matrix},
\\
&(2)
&
\lim_{\nu\rightarrow \infty}
 N_\nu^{-1} \sum_{1 \leq k \neq l \leq N_\nu}
    \|W_{\nu k}\| \|W_{\nu l}\|  \cov( D_{\nu k}, D_{\nu l})
     = 0,  \quad\mbox{and}
\\
&(3)
 &\lim_{\nu\rightarrow \infty}
 N_\nu^{-1} \sum_{k=1}^{N_\nu}  \|W_{\nu k} \|^2 
P( \|W_{\nu k}\|  |D_{\nu k} - \pi_{\nu k} |\geq \varepsilon  \sqrt{N_\nu}  )
=  0 \  \mbox{for any}\  \varepsilon>0.
\eas
Then, as $\nu\rightarrow\infty,$
\(
S_\nu^{-1} N_\nu^{-1/2} \sum_{k=1}^{N_\nu} W_{\nu k}  ( D_{\nu k} - \pi_{\nu k})
\convergeto N(0,  I_{q+1}).
\)

\end{theorem}

For notational simplicity,
we shall suppress the subscript $\nu$
and use $N\rightarrow \infty$ instead of   $\nu \rightarrow \infty$.
This theorem implies that
\(
 N^{-1/2} \sum_{k=1}^{N } W_{ k}  ( D_{  k} - \pi_{  k})
\convergeto N(0,  V_0)
\).
This result is sufficient to establish the asymptotic normalities of the IPW and ELW estimators.
Define   $B_2 = \lim_{N \rightarrow\infty}  (1/N)\sum_{k=1}^N \{g(Z_k)\}^{\otimes 2}$.
We redefine the notation  from Section \ref{sec2}.
For two generic functions $f$ and $h$,  define
$
B_{fh} =  \lim_{N \rightarrow\infty}   N^{-1}  \sum_{k=1}^N  f(Z_k) h^\T (Z_k) /\pi_k
$
and write   $B_{11} = \lim_{N \rightarrow\infty}  N^{-1} \sum_{k=1}^N  1 /\pi_k    $ and
$B_{f1} = \lim_{N \rightarrow\infty}  N^{-1}  \sum_{k=1}^N  f(Z_k)  /\pi_k $.

\begin{theorem}
\label{CLT-B-ups2}
Under the conditions in Theorem \ref{CLT-B-ups},   as $\nu$ goes to infinity,
\ben
\item[(a)]
\(
  N^{1/2} (\hat \theta_{\ipw} - \theta_0)
 \convergeto N(0,  \Sigma_{\ipw}),
\)
where
$
\Sigma_{\ipw}
=   B_{gg} - B_2;
$

\item[(b)]
$
\sqrt{N_{\nu}} (\hat \theta_{\sipw}  - \theta_{  0})
\convergeto N(0,  \Sigma_{\sipw}),
$
where
\[
\Sigma_{\sipw}
=
\Sigma_{\ipw} -
   \frac{(B_{g1} - \theta_{ 0})^{\otimes 2}}{ B_{11}-1 }
   + \frac{ (B_{g1} - \theta_{0} B_{11})^{\otimes 2} }{ B_{11}-1 };
  \]

\item[(c)]
\(
\sqrt{N_\nu}(\hat \theta_{\elw} - \theta_0)
\convergeto N(0, \Sigma_{\elw}),
\)
where
\(
\Sigma_{\elw} =
\Sigma_{\ipw} -
 (B_{g1} - \theta_{ 0})^{\otimes 2} /( B_{11}-1 );
\)

\item[(d)]
the ELW estimator $\hat \theta_{\elw}$
is more efficient than  $\hat \theta_{\ipw}$
and   $\hat \theta_{\sipw},$
i.e.
\(
  \Sigma_{\elw} \leq \Sigma_{\ipw}
\)
and
\(
  \Sigma_{\elw} \leq \Sigma_{\sipw}
,\)
where the equalities hold only if the $\pi_{k}$ are all equal.

\een

\end{theorem}

By Theorem \ref{CLT-B-ups2},  again    the ELW,  IPW, and   SIPW estimators are asymptotically normal,
and the ELW estimator is asymptotically the most efficient.
A variance estimator similar to \eqref{Sigma-ELW-est} can be constructed as
\(
\widehat\Sigma_{\elw} =
 \hat B_{gg} - \hat B_2   -
 (\hat B_{g1} - \hat \theta_{\elw})^{\otimes 2}/(\hat  B_{11}-1),
\)
where
$
\hat B_{fh} =    N   \sum_{k=1}^N  (\hat p_k)^2 f(Z_k) h^\T (Z_k)
$
and
$
\hat B_2 =    \sum_{k=1}^N   \hat p_k  \{ g(Z_k)\}^{\otimes 2}.
$

\begin{remark}
Under UPS-WOR, the
 $Z_{\nu k}$ are non-random and
 only  the $D_{\nu k}$
are random variables.
Unlike the cases in  missing data problems,
the ELW weights $\hat p_{\nu k}$  do not
lead to an estimate of  a certain  distribution function.
They are simply taken as  weights  for parameter estimation.
\end{remark}

\subsection{ELW for UPS-WR}

We still assume that the finite population of interest $\{Z_1, \ldots, Z_N\}$,
where  the subscript $\nu$  is suppressed,
lies in a hypothetical infinite sequence
(indexed by $\nu$) of finite populations
of increasing size.
The parameter to be estimated is $\theta_{0} = (1/N)\sum_{i=1}^N g(Z_i)$.
Suppose a sample of size $n$ is drawn   by an UPS-WR from
$\{Z_1, \ldots, Z_N\}$.
Let the sampling probability of the $i$th individual
at each sampling occasion be $ q_i^*$ ($i=1, 2, \ldots, N$).
Clearly, $q_i^*>0$ and $\sum_{i=1}^N q_i^*  =1$.
Denote the observed data by  $z_i$
and the corresponding sampling probability by $q_i$
for $i=1, 2, \ldots, n$.

To apply the IPW and ELW methods from Section \ref{sec2}, we  define $\pi_i^* = nq_i^*$ ($1\leq i\leq N$)
 and  $\pi_i = nq_i$ ($1\leq i\leq n$), which may be greater than 1.
 The  IPW estimator of $\theta_{  0} $ under UPW-WR is the
 Hansen--Hurwitz estimator \citep{Hansen1943},
\ba
\label{eq-HH}
\hat \theta_{\ipw}  =  \frac{1}{nN} \sum_{i=1}^n  \frac{g(z_i )}{q_i}
=  \frac{1}{N} \sum_{i=1}^n  \frac{g(z_i )}{\pi_i}.
\ea

The ELW method from Section \ref{sec2} was introduced with
data $\{ (D_i, D_iZ_i, \pi(Z_i)): 1\leq i\leq N \}$.
However, the notation  $D_i$  is not appropriate under UPS-WR, because
a unit from the population may be drawn multiple times.
As the ELW estimator depends only on
$N$ and   $\{  (z_i, \pi_i): 1\leq i\leq n \}$,
 it is applicable to UPS-WR  through a modified version
 of Algorithm \ref{elw-procedure}.

\begin{algorithm}[h]
\caption{ELW estimation procedure under UPS-WR  \label{elw-procedure-ups-wr}}

\renewcommand{\algorithmicrequire}{\textbf{Input:}} 
\renewcommand{\algorithmicensure}{\textbf{Output:}} 
\begin{algorithmic}

\REQUIRE
The sample drawn, $\{ (  g(z_i), \pi_i): i=1, 2, \ldots, n \}$;
The size of the finite population, $N$.

\ENSURE
The ELW estimate, $\hat \theta_{\elw}$, of $\theta = \e\{g(Z)\}$.

\STATE   {\bf Step 1.}
Calculate
\(
\xi_i =  n /N +(1-n /N) \pi_i
\) ($1\leq i \leq n$),
 $\zeta_l = \min_{1\leq i\leq n}\pi_i$,
 and $\zeta_u = \min_{1\leq i\leq n}\xi_i$.

\STATE  {\bf Step 2.}
Calculate  $\hat \alpha$   by solving
$
0 =
  \sum_{i=1}^{n }   (\pi_i - \alpha )/(  \xi_i  - \alpha )
$
 in the interval $[\zeta_l, \zeta_u)$,
 and calculate $  \lambda(\hat \alpha)   =  ( N - n )/\{ n (1- \hat \alpha) \}$.

\STATE  {\bf Step 3.}
Calculate
$
\hat p_i  =  n^{-1}  \{1+ \lambda (\hat \alpha) (\pi_i - \hat \alpha ) \}^{-1},
i=1, 2, \ldots, n$.

\STATE  {\bf Step 4.}
Calculate the ELW estimate
$
\hat \theta_{\elw} =  \sum_{i=1}^n \hat p_i  g(z_i)
$.

\end{algorithmic}
\end{algorithm}

Define $\alpha_{0} = n/N$
and
$
B_{fh} =    N^{-1}  \sum_{k=1}^N  f(Z_k) h^\T (Z_k) /\pi_k^*
$
for two generic functions $f$ and $h$.
The asymptotic normalities of the IPW, SIPW, and ELW estimators
are established in the following theorem.

\begin{theorem}
 \label{CLT-C-ups}
Suppose that there exist $\alpha_0^*>0,$ $\theta_0^*,$ $B_{11}^*> 1/\alpha_0^*,$ $B_{g1}^*,$
and  a positive-definite matrix $B_{gg}^*$ such that
as $\nu\rightarrow\infty,$  we have
$\alpha_0 \rightarrow \alpha_0^*,$
$\theta_{ 0}\rightarrow \theta_0^*,$
$B_{1}  \rightarrow B_{11}^*,$
$B_{g1}  \rightarrow B_{g1}^*,$
and
$B_{gg}  \rightarrow B_{gg}^*$.
 As $\nu\rightarrow\infty,$
\ben
\item[(a)]
$
\sqrt{n}(\hat \theta_{\ipw} - \theta_0)
 \convergeto N(0,  \Sigma_{\ipw})
$
with
 $ \Sigma_{\ipw} =  \alpha_{0}^* B_{gg}^* - (\theta_0^*)^{\otimes 2};$
\item[(b)]
$
\sqrt{n} (\hat\theta_{\sipw} - \theta_0)
 \convergeto   N(0, \Sigma_{\sipw }),
$
with
$
\Sigma_{\sipw }
=
  \alpha_{0}^* \{
  B_{gg}^* - \theta_0^* B_{g1}^* - B_{g1}^*\theta_0^* + (\theta_0^*)^{\otimes 2} B_{11}^*\}
;$

\item[(c)]
$
\sqrt{n}(\hat \theta_{\elw} - \theta_0)
 \convergeto N(0,  \Sigma_{\elw}),
$
where
 \bas
\Sigma_{\elw}
&=&
   \alpha_{0}^*B_{gg}^*
   -  (\theta_{0}^*)^{\otimes 2}
   + \frac{ (1- \alpha_{0}^* )^2 (  \theta_{0}^*B_{11}^* - B_{g1}^*)^{\otimes 2} }{
     (\alpha_{0}^* B_{11}^* - 1)(B_{11}^* - 1)^2  }
   -  \frac{ (\alpha_{0}^*B_{g1}^*- \theta_{0}^* )^{\otimes 2}}{ \alpha_{0}^* B_{11}^* - 1  }.
 \eas
\een
\end{theorem}

As $\hat p_i \approx 1/(N\pi_i)$  ($1\leq i\leq n$),
similarly to \eqref{Sigma-ELW-est}, we propose to estimate $\Sigma_{\elw}$ by
\[
\widehat\Sigma_{\elw} =
 \alpha_{0} \hat B_{gg}
   -  (\hat \theta_{\elw} )^{\otimes 2}
   + \frac{ (1- \alpha_{0}  )^2 (  \hat \theta_{\elw}\hat  B_{11}  - \hat  B_{g1} )^{\otimes 2} }{
     (\alpha_{0} \hat  B_{11}  - 1)(\hat B_{11}  - 1)^2  }
   -  \frac{ (\alpha_{0} \hat  B_{g1} -  \hat  \theta_{\elw}  )^{\otimes 2}}{ \alpha_{0} \hat  B_{11} - 1  },
\]
where
$
\hat B_{fh} =    N   \sum_{k=1}^n  (\hat p_k)^2 f(z_k) h^\T (z_k).
$

\section{Simulation study}\label{sec4}

We conduct simulations to investigate the finite-sample performance
of the proposed ELW estimator and the accompanying
asymptotic-normality-based interval estimator.
For comparison, we also take into account  the IPW  estimator, the SIPW estimator,
and some popular variants of the IPW estimator:
\ben
\item
The modified IPW estimator of \cite{Zong2019} (ZZZ for short):
\(
\hat \theta_\zzz =  N^{-1} \sum_{i=1}^N    D_iY_i / \tilde \pi_i,
\)
where $\tilde \pi_i =  \max\{\pi_{(K)}, \pi(X_i) \}$,
$K$ is the maximum $i$ such that $\pi_{(i)}\leq 1/(i+1)$,
and  $ \{ \pi_{(1)},\ldots, \pi_{(N)} \} $   are the propensity scores  in increasing order.
\item
The  trimmed IPW estimator of \cite{Crump2009} (CHIM for short):
 \bas
\hat \theta_\chim = \left. \sum_{i=1}^N   \frac{ D_iY_i }{ \pi(X_i)}\cdot I\{  \alpha \leq \pi(X_i)\leq 1-  \alpha \}\right/
 \sum_{i=1}^N   I\{  \alpha \leq \pi(X_i)\leq 1-  \alpha \},
\eas
where $ \alpha$ is obtained by minimizing a variance term and $I(\cdot)$
is the indicator function.

\item
The IPW estimator of \cite{Ma2020}  with $s=1$  and $s=2$, denoted by MW1 and MW2, respectively.
Following  \cite{Ma2020}, we set
the tuning parameters $b_N$  and $h_N$ in MW1 and MW2  to  the respective solutions of
$
b_N^s N^{-1} \sum_{i=1}^N  I\{ \pi(X_i)\leq b_N \} = 1/(2N)
$
and
$
  h_N^5  \sum_{i=1}^N  I\{ \pi(X_i)\leq h_N \} =1.
$
For details, see    the discussion below  Theorem 3 of \cite{Ma2020}
and Section III of their supplementary material.
\een

We simulate data from two examples, which represent missing data
and unequal probability sampling problems, respectively.
All numbers reported in this   simulation study are calculated
based on $M=5000$ simulated random samples.

\begin{example}[Missing data]
\label{ex1}
Instead of generating $X,$
we  generate the propensity score $\pi(X)$
 from  $P(\pi(X)\leq u) =  u^{\gamma - 1}$ $(0\leq u\leq 1)$
with $\gamma = 1.5 $ or $  2.5$.
Given $\pi(X),$
we generate  $Y$ from
$Y = \mu\{ \pi(X) \} + c\cdot (\eta-4)/\sqrt{8},$
where $c=1$ or $0.1,$ and $\eta \sim \chi_4^2,$
and the missingness status $D$ of $Y$
follows the Bernoulli distribution with success probability $\pi(X)$.
Four choices of $\mu(t)$ are considered:
$\mu(t) = \cos(2\pi t)$ (Model 1),
 $\mu(t) = 1 - t $ (Model 2),
  $\mu(t)  =   \cos(2\pi t) + 5$ (Model 3),
  and $\mu(t)  =   6-t$ (Model 4).
  The sample size is $N =2000$ and the parameter of interest is $\theta = \e(Y)$.
\end{example}

This example is a modified version of Example 1 in  Section III
of the supplementary material of \cite{Ma2020},
who  considered the cases with $\gamma=1.5$, $c=1$, and $N=2000$ for Models 1 and 2.
The parameter $\gamma  $ ($\gamma>1$) controls the tail behaviour of $1/\pi(X)$.
When $\gamma > 2$, the tail is light and
$\e\{ 1/\pi(X)\} = (\gamma-1)/(\gamma-2)$ is finite.
In this case, if $g$ is bounded,  then the conditions in Theorem \ref{CLT-A-known}
are generally fulfilled,  and the asymptotic normalities of the ELW, IPW, and SIPW
estimators are guaranteed.
However, in the case of $1<\gamma\leq 2$,
the tail is heavy and  $\e\{ 1/\pi(X)\} = \infty$,
which violates the conditions of Theorem \ref{CLT-A-known}:
the ELW, IPW, and SIPW estimators no longer follow asymptotically normal distributions.
The constant $c$ controls the influence of the random error on the
response variable; a smaller $c$ leads to a smaller noise.
Models 3 and 4 are simply Models 1 and 2 with a mean shift.

\begin{figure}
\centering
\mbox{
\includegraphics[width=0.43\textwidth, height=0.23\textheight]{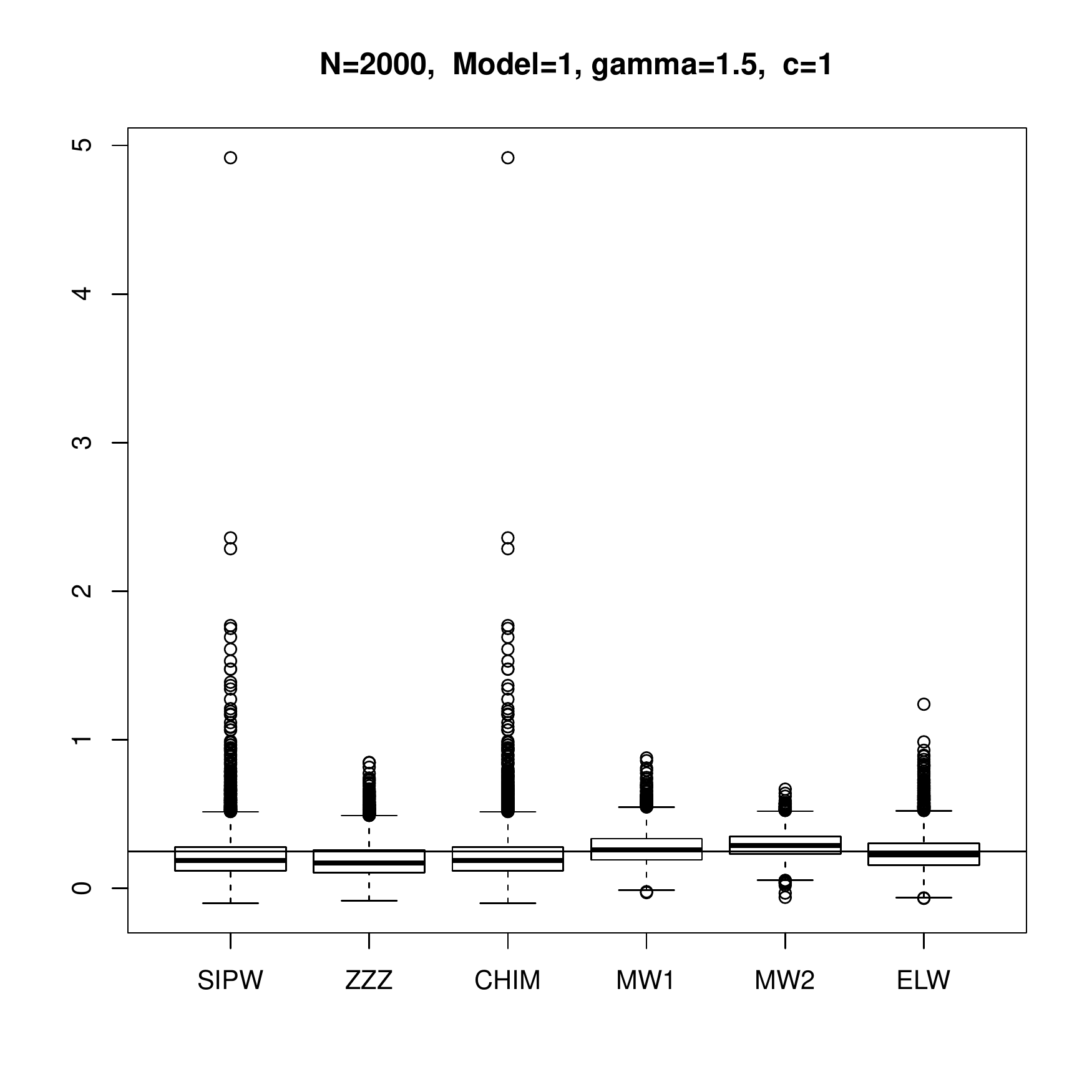}
\includegraphics[width=0.43\textwidth, height=0.23\textheight]{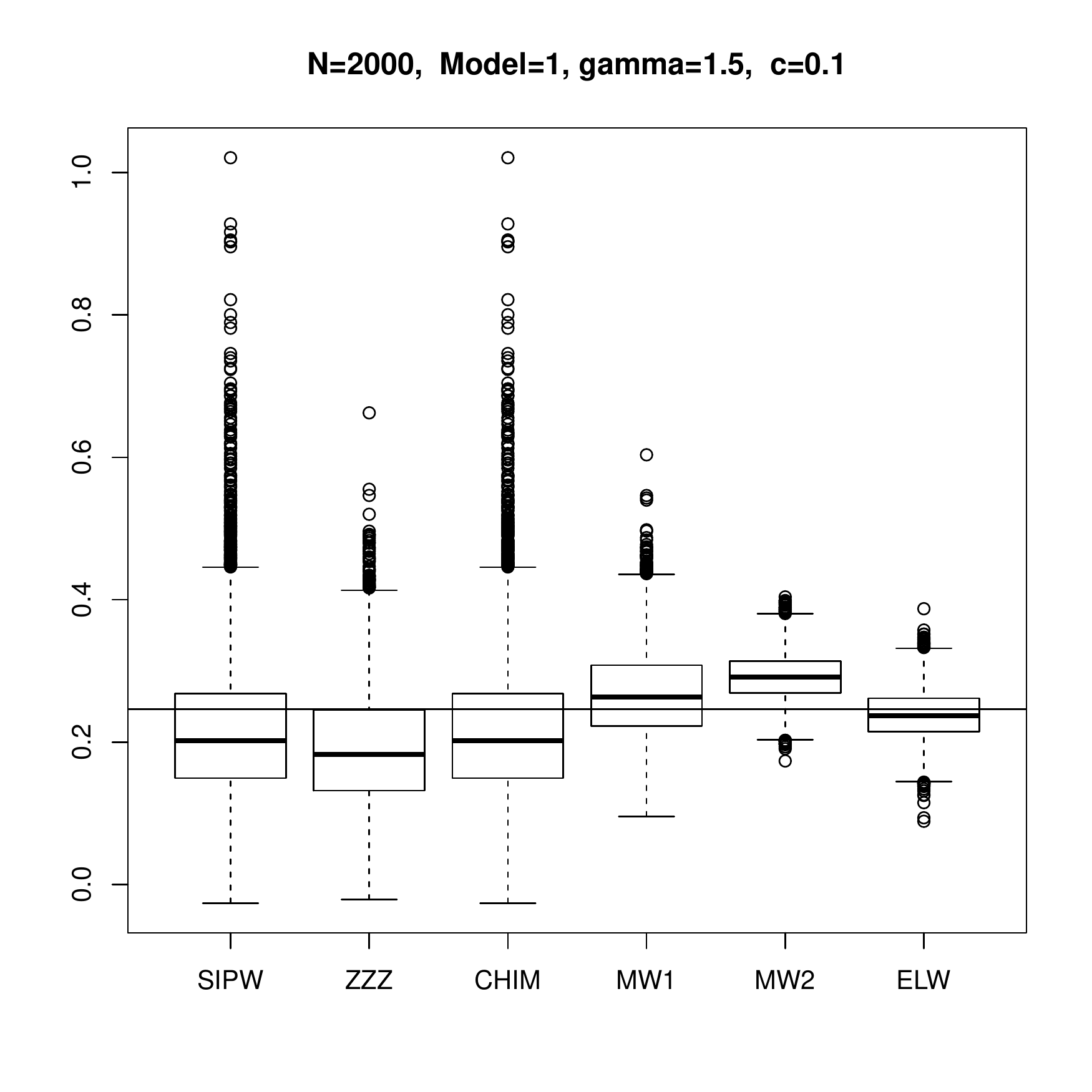}  }\\
\mbox{
\includegraphics[width=0.43\textwidth, height=0.23\textheight]{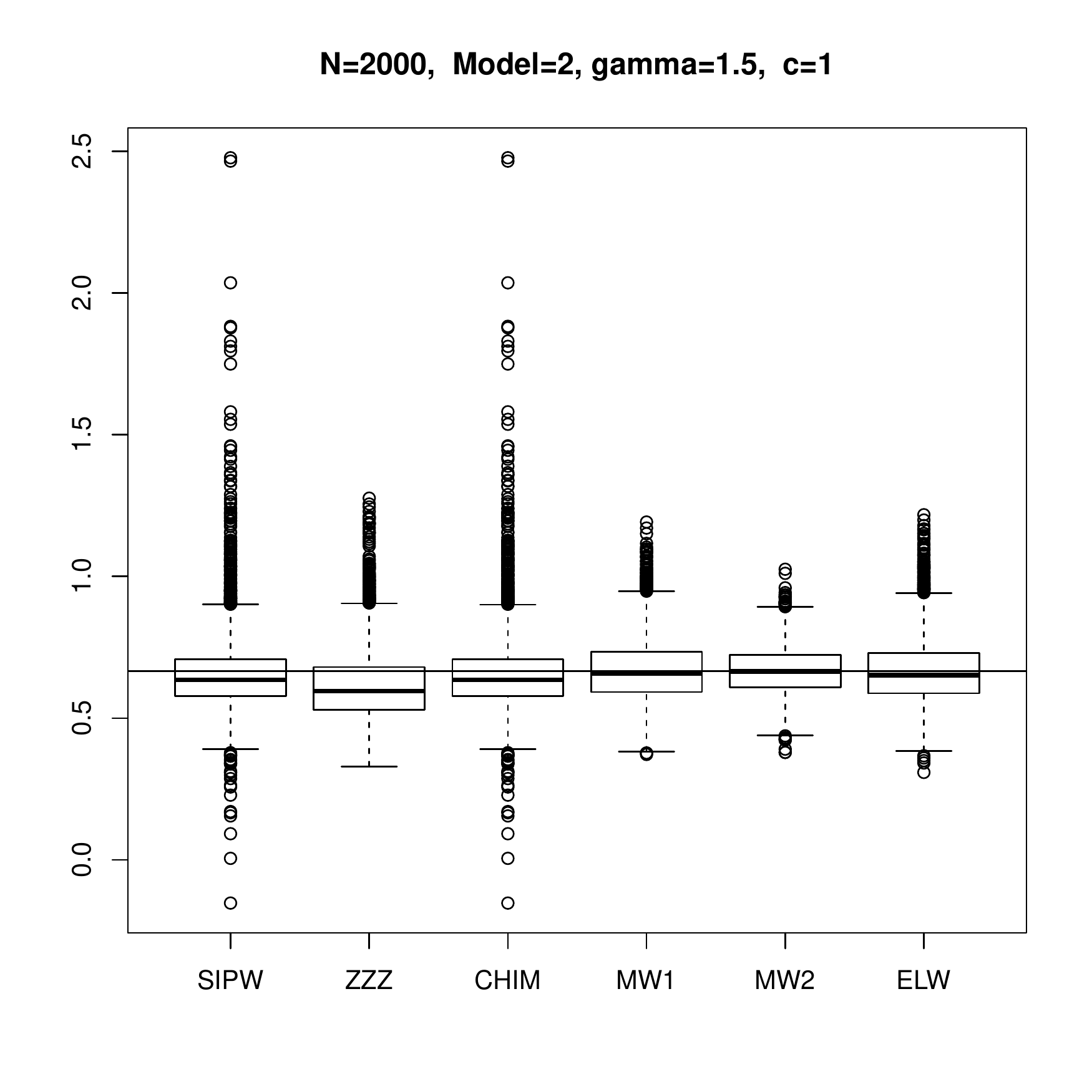}
\includegraphics[width=0.43\textwidth, height=0.23\textheight]{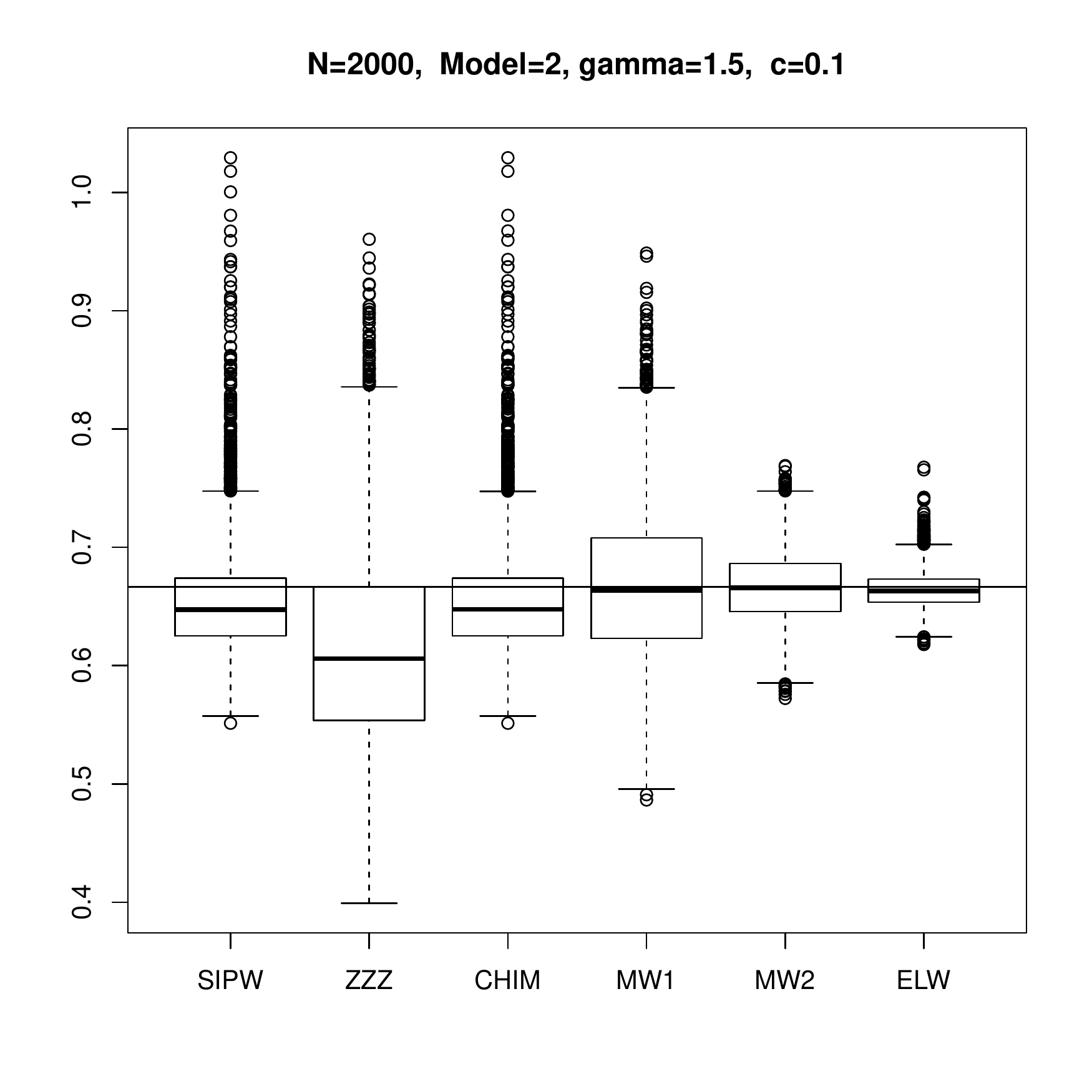}}\\
\mbox{
\includegraphics[width=0.43\textwidth, height=0.23\textheight]{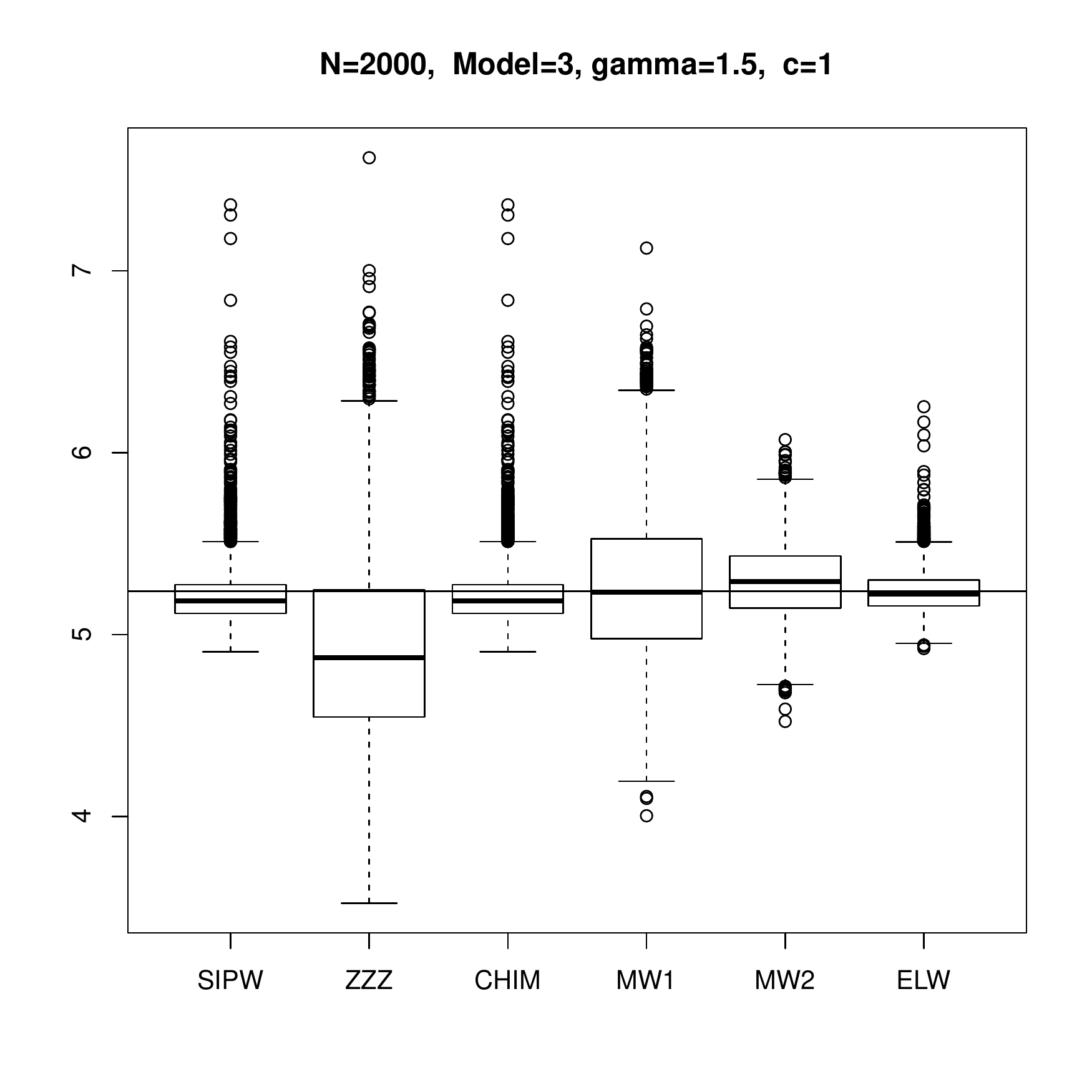}
\includegraphics[width=0.43\textwidth, height=0.23\textheight]{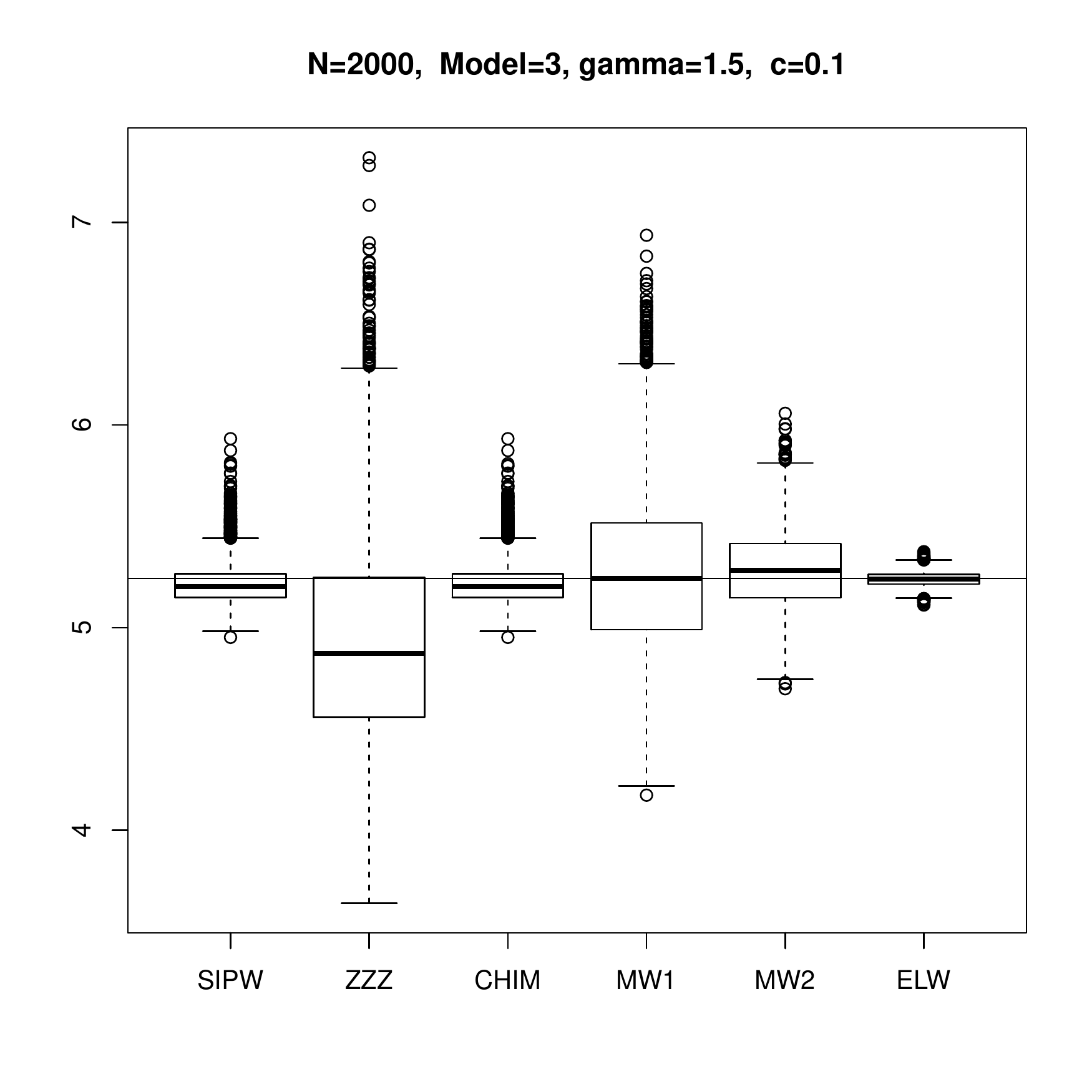} }\\
\mbox{
\includegraphics[width=0.43\textwidth, height=0.23\textheight]{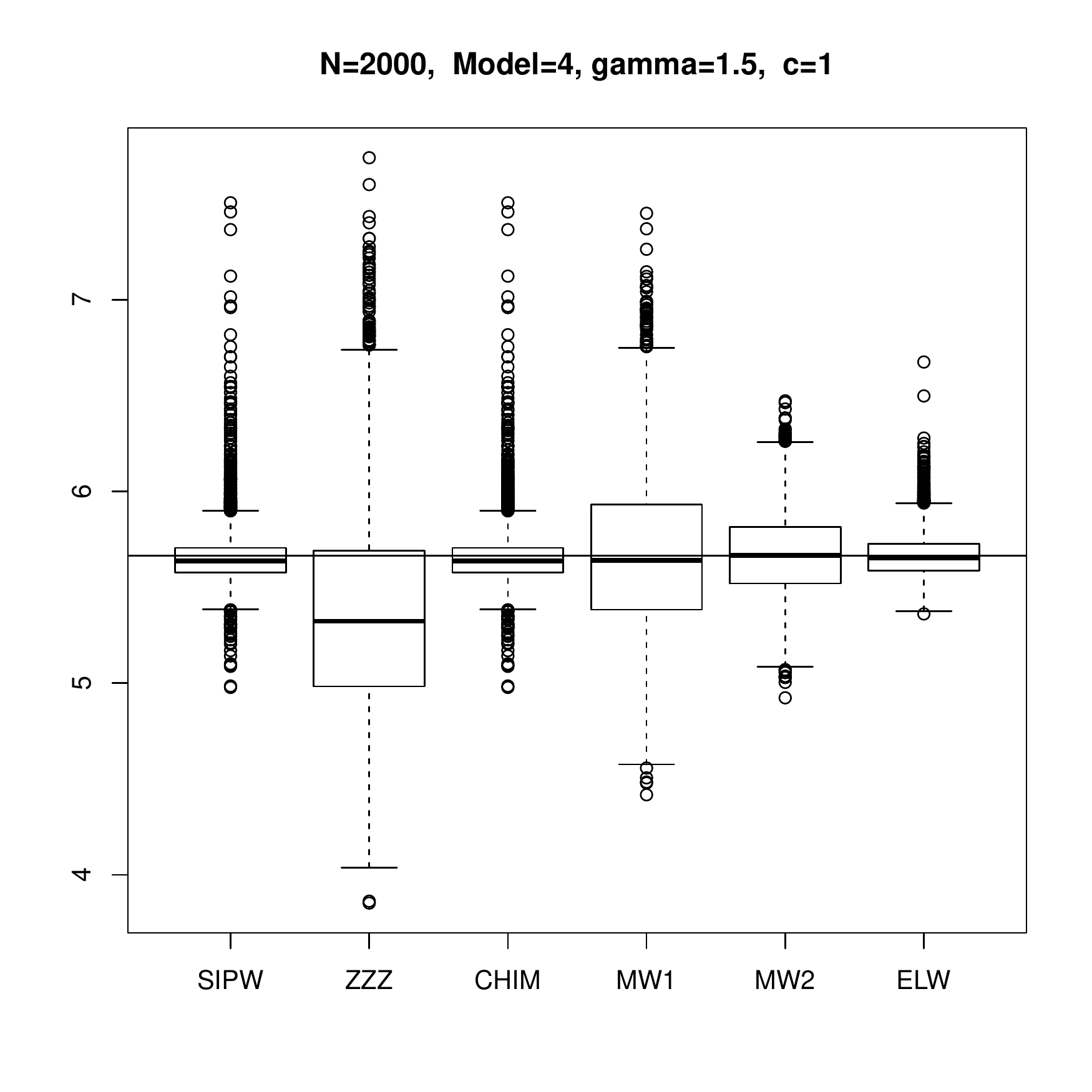}
\includegraphics[width=0.43\textwidth, height=0.23\textheight]{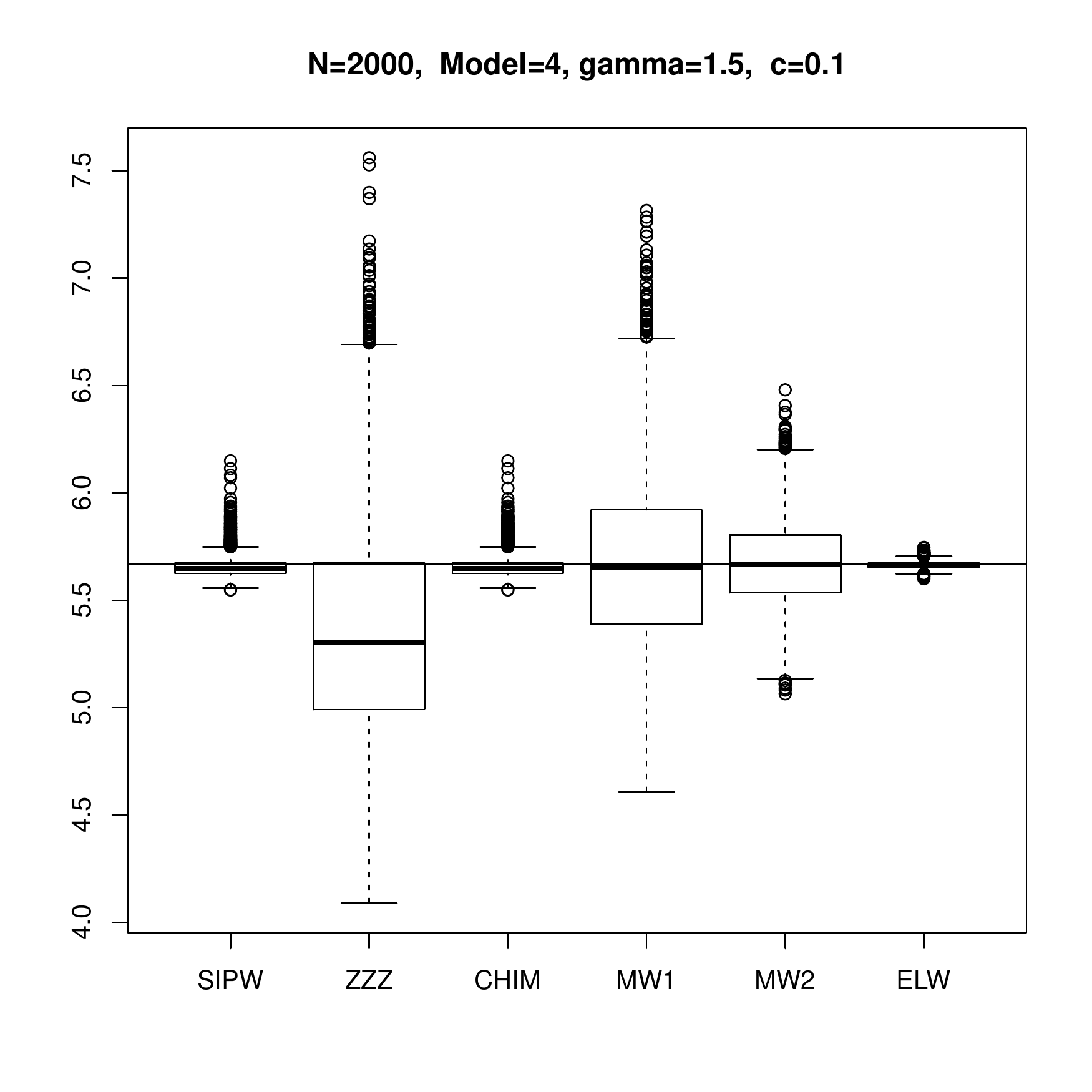} }
\caption{Boxplots of the SIPW, ZZZ, CHIM, MW1, MW2, and ELW estimators
when data were generated from  Example \ref{ex1} with  $N=2000$ and $\gamma=1.5$.
The solid horizontal line corresponds to the target parameter value. }
\label{fig-ex1-gamma=1.5}
\end{figure}

\begin{figure}
\centering
\mbox{
\includegraphics[width=0.43\textwidth, height=0.23\textheight]{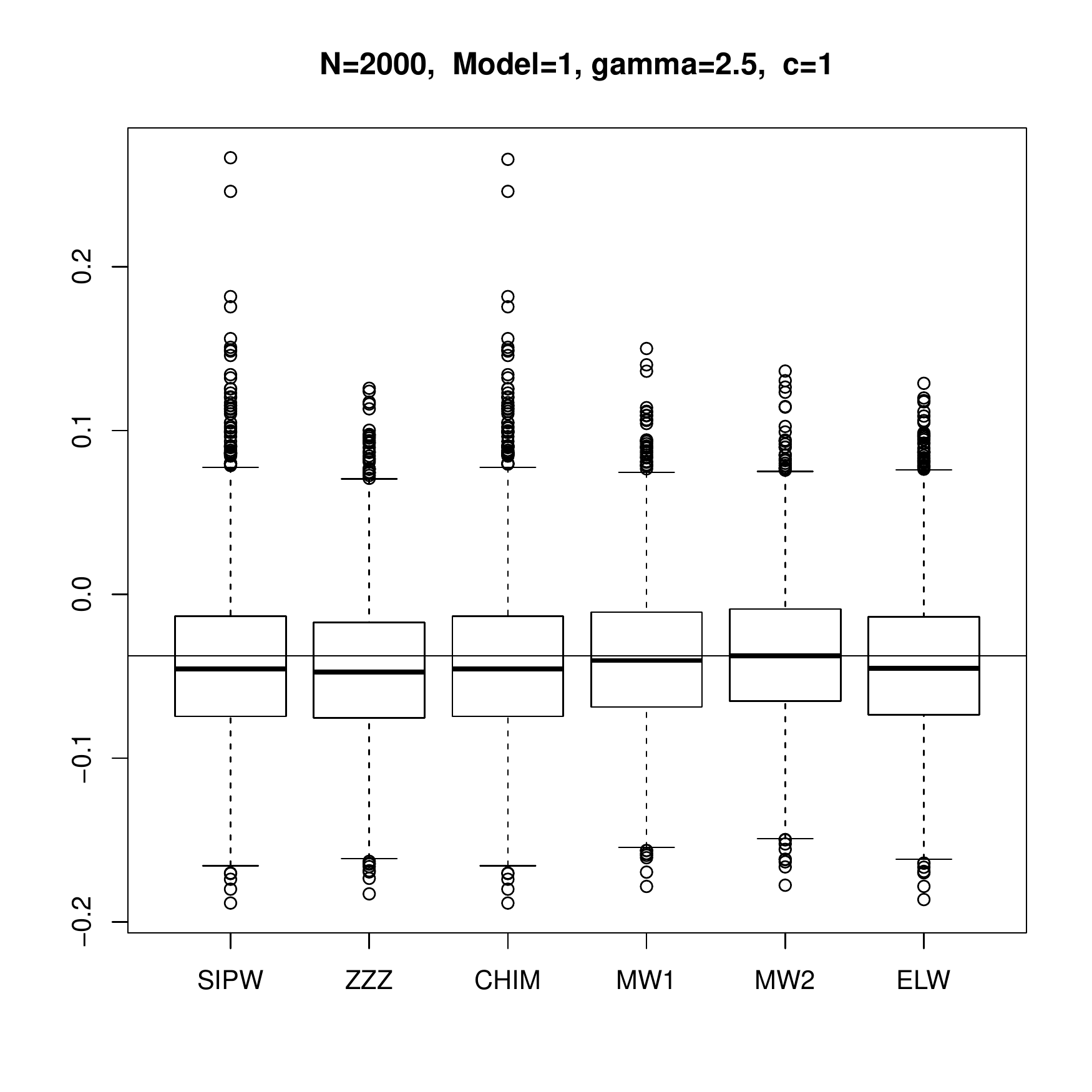}
\includegraphics[width=0.43\textwidth, height=0.23\textheight]{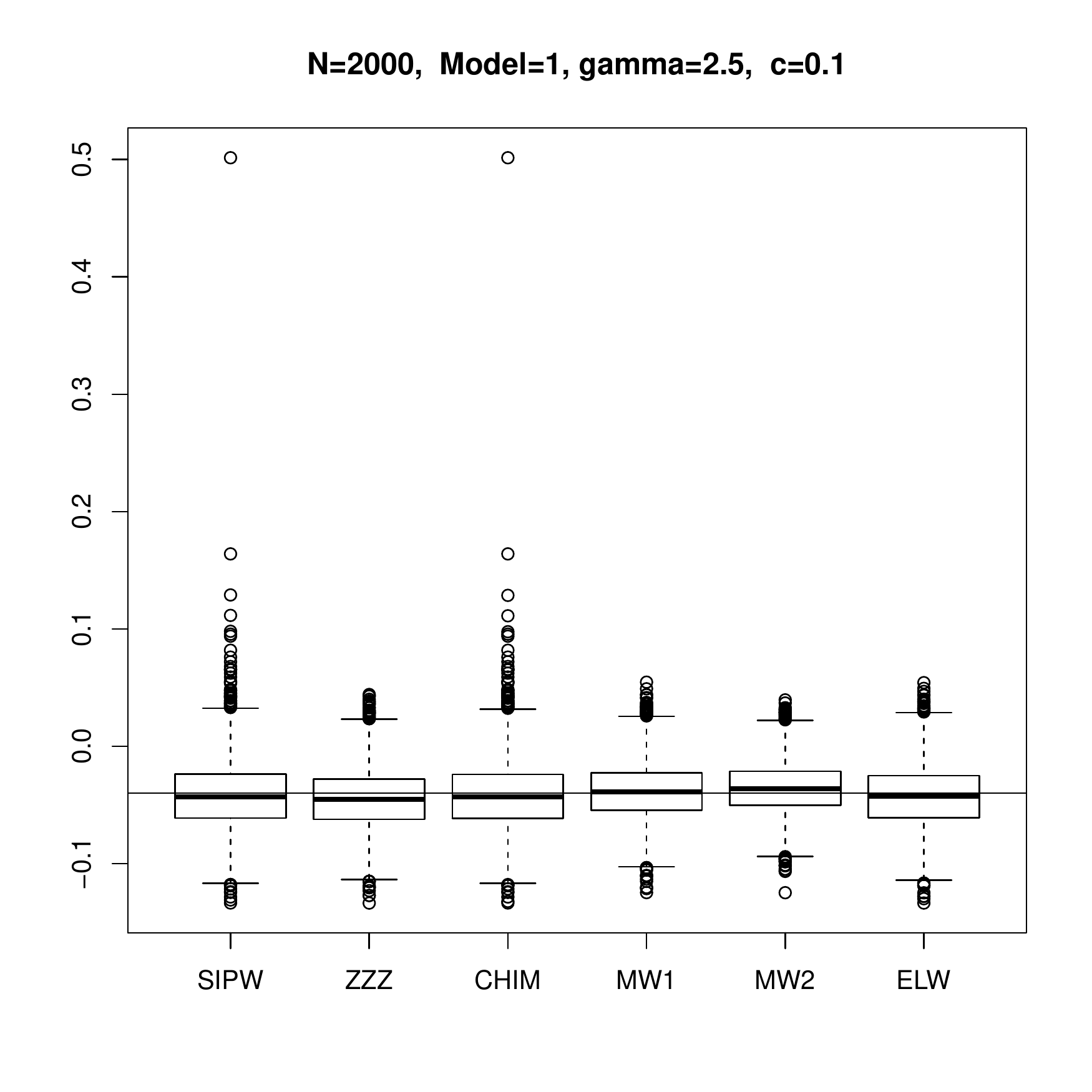} }\\
\mbox{
\includegraphics[width=0.43\textwidth, height=0.23\textheight]{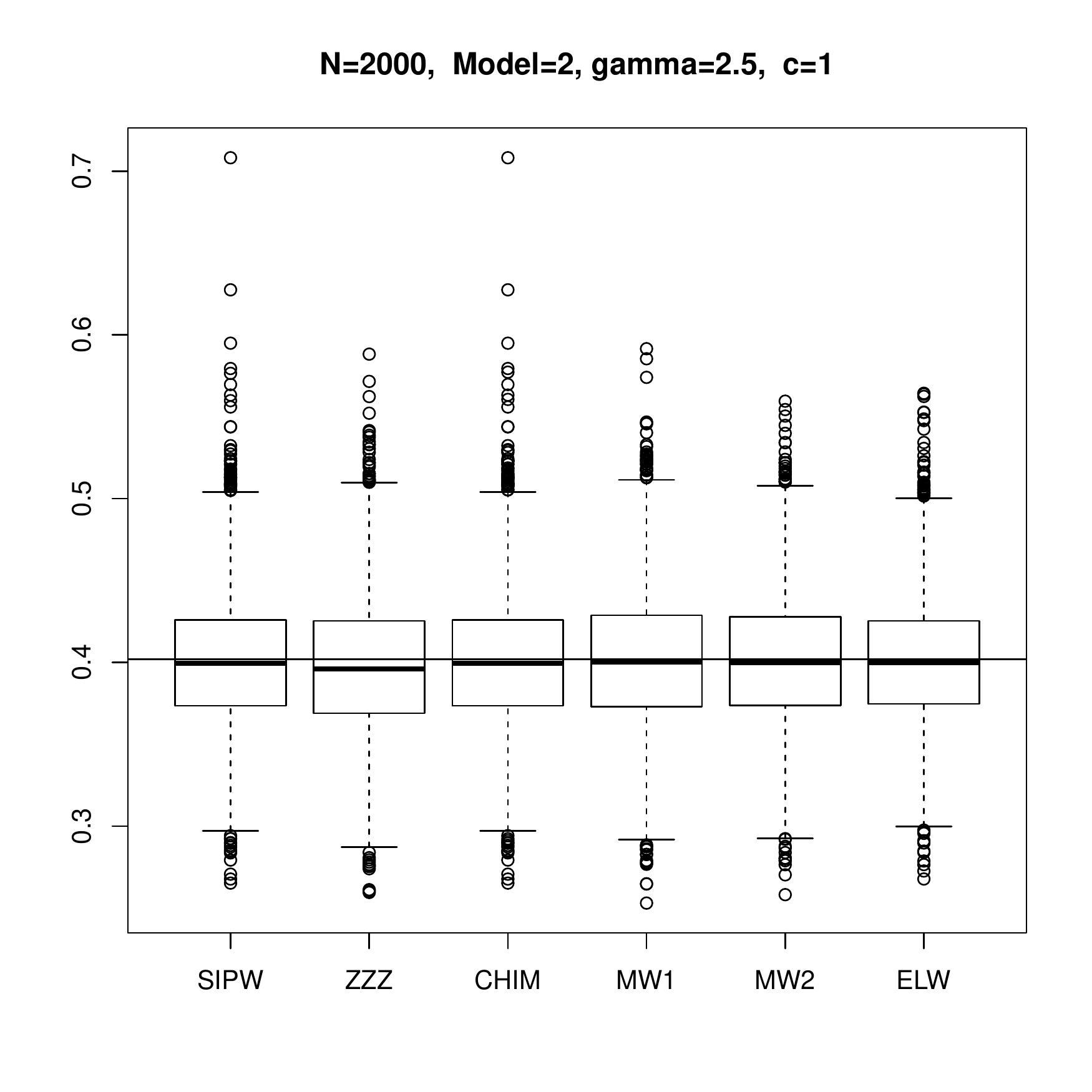}
\includegraphics[width=0.43\textwidth, height=0.23\textheight]{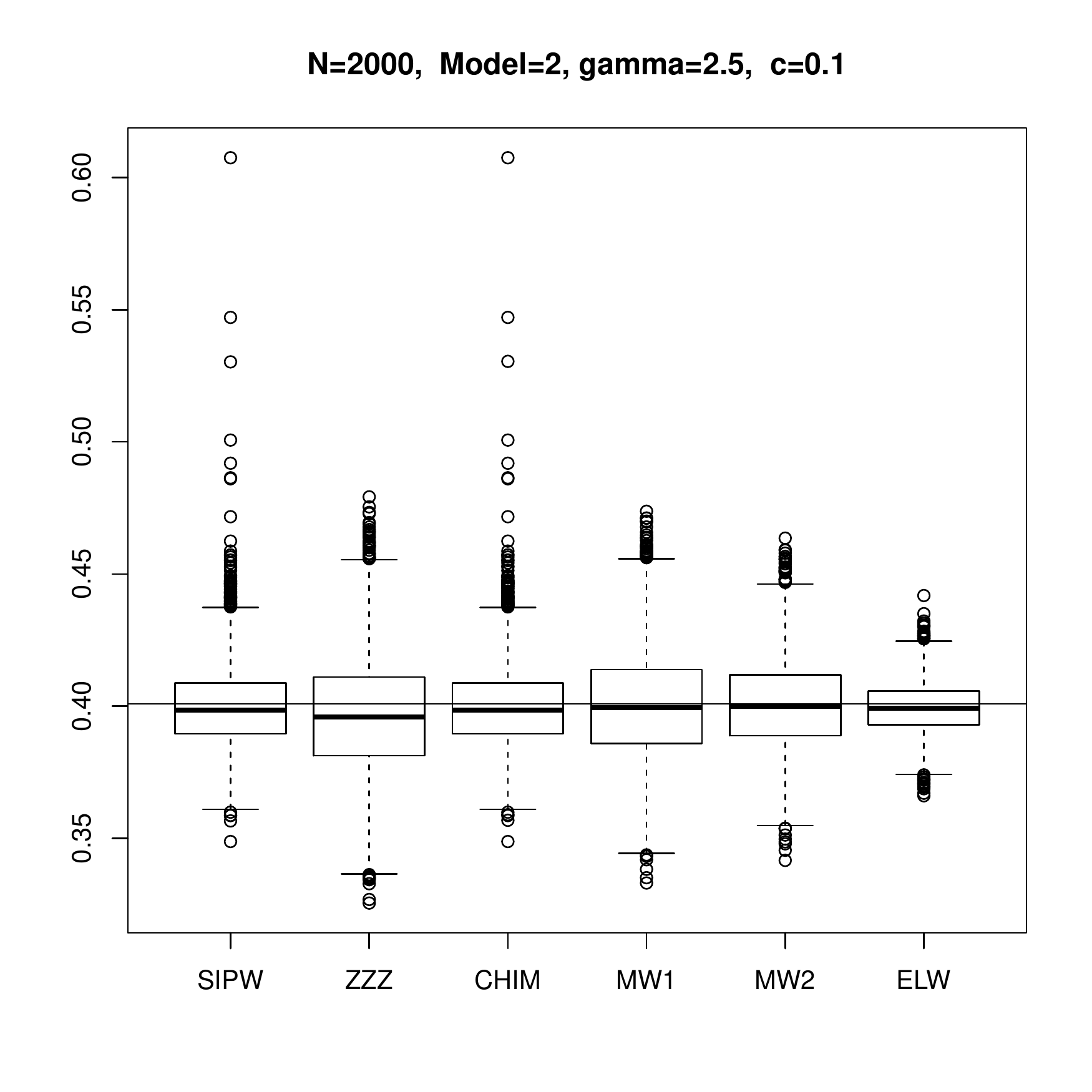} }\\
\mbox{
\includegraphics[width=0.43\textwidth, height=0.23\textheight]{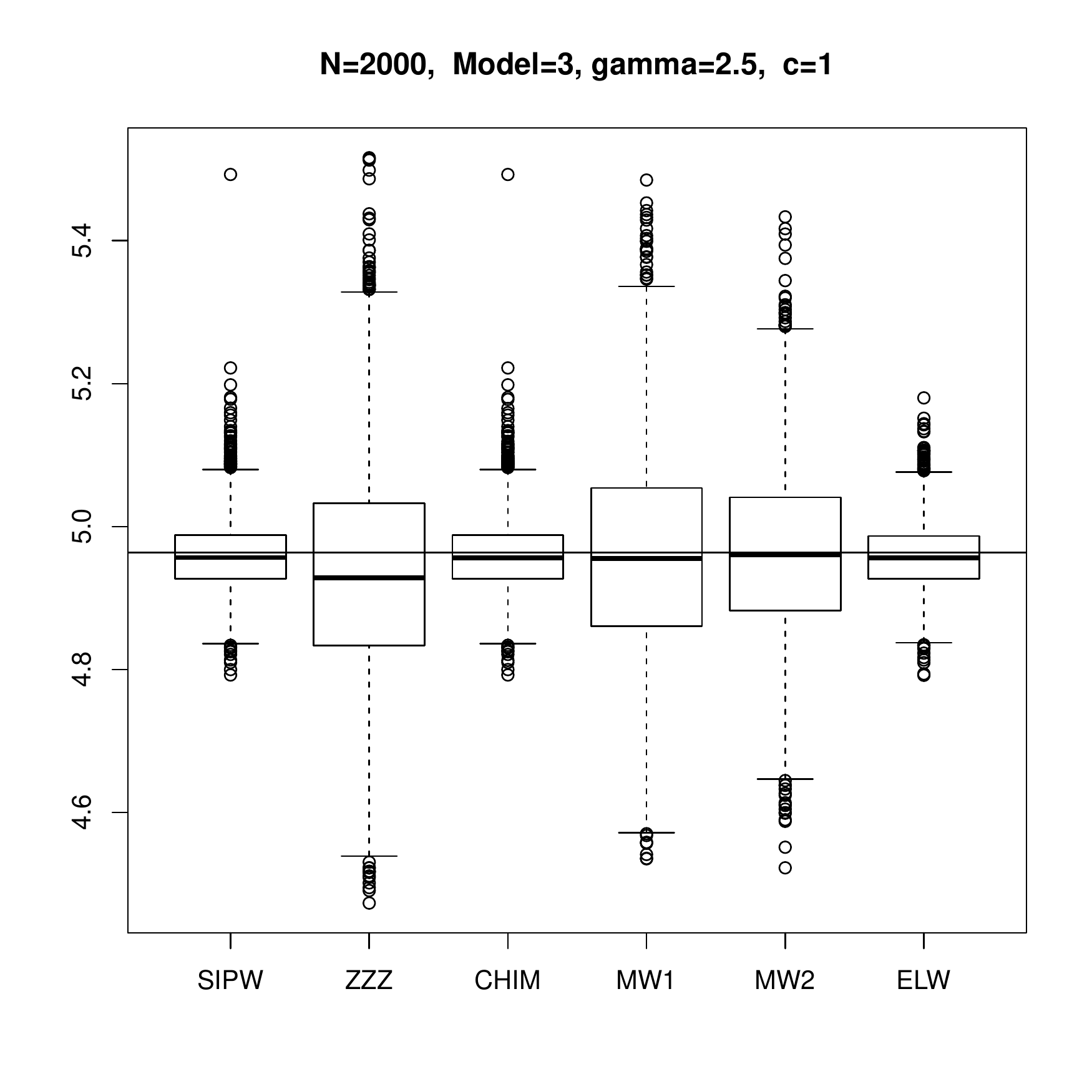}
\includegraphics[width=0.43\textwidth, height=0.23\textheight]{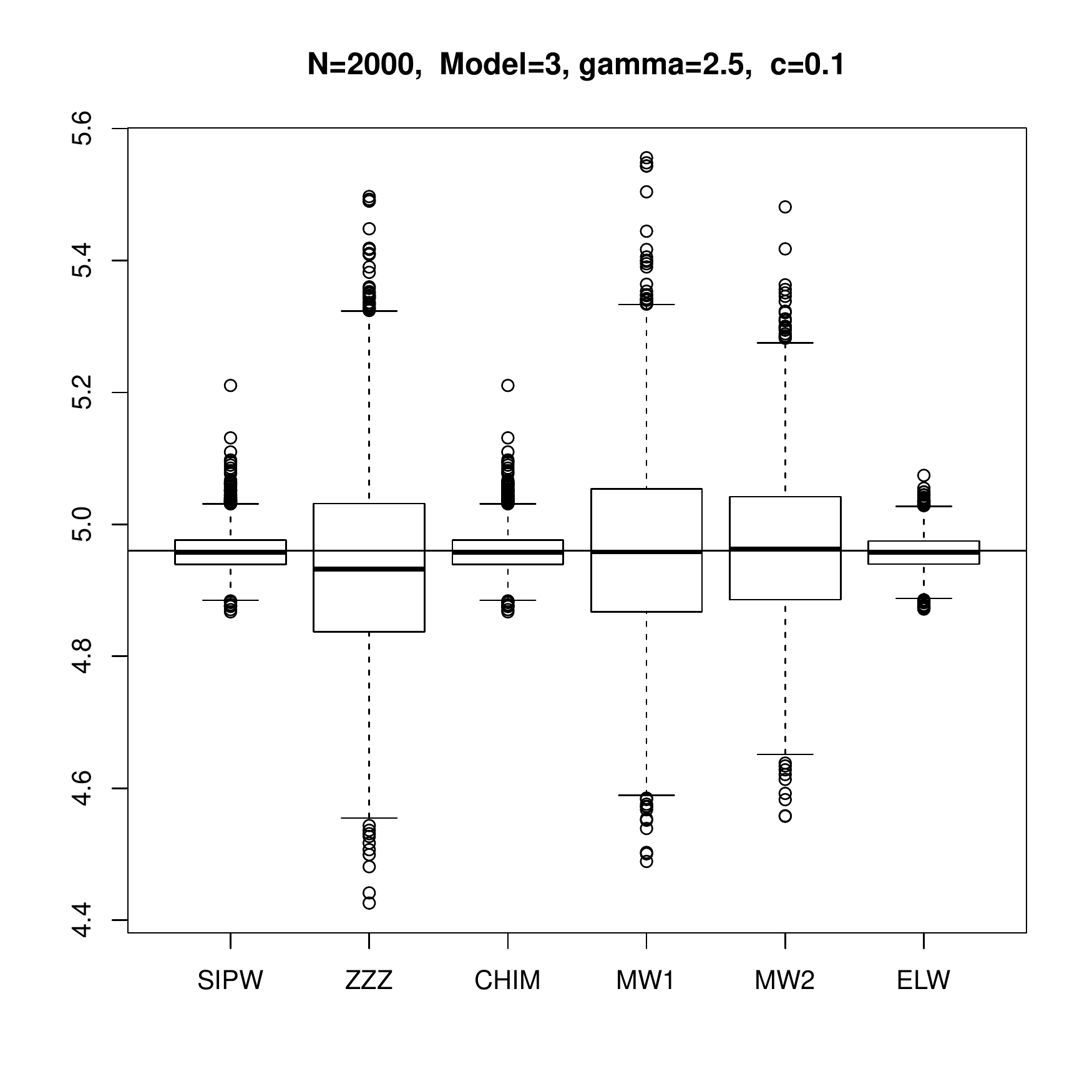} }\\
\mbox{
\includegraphics[width=0.43\textwidth, height=0.23\textheight]{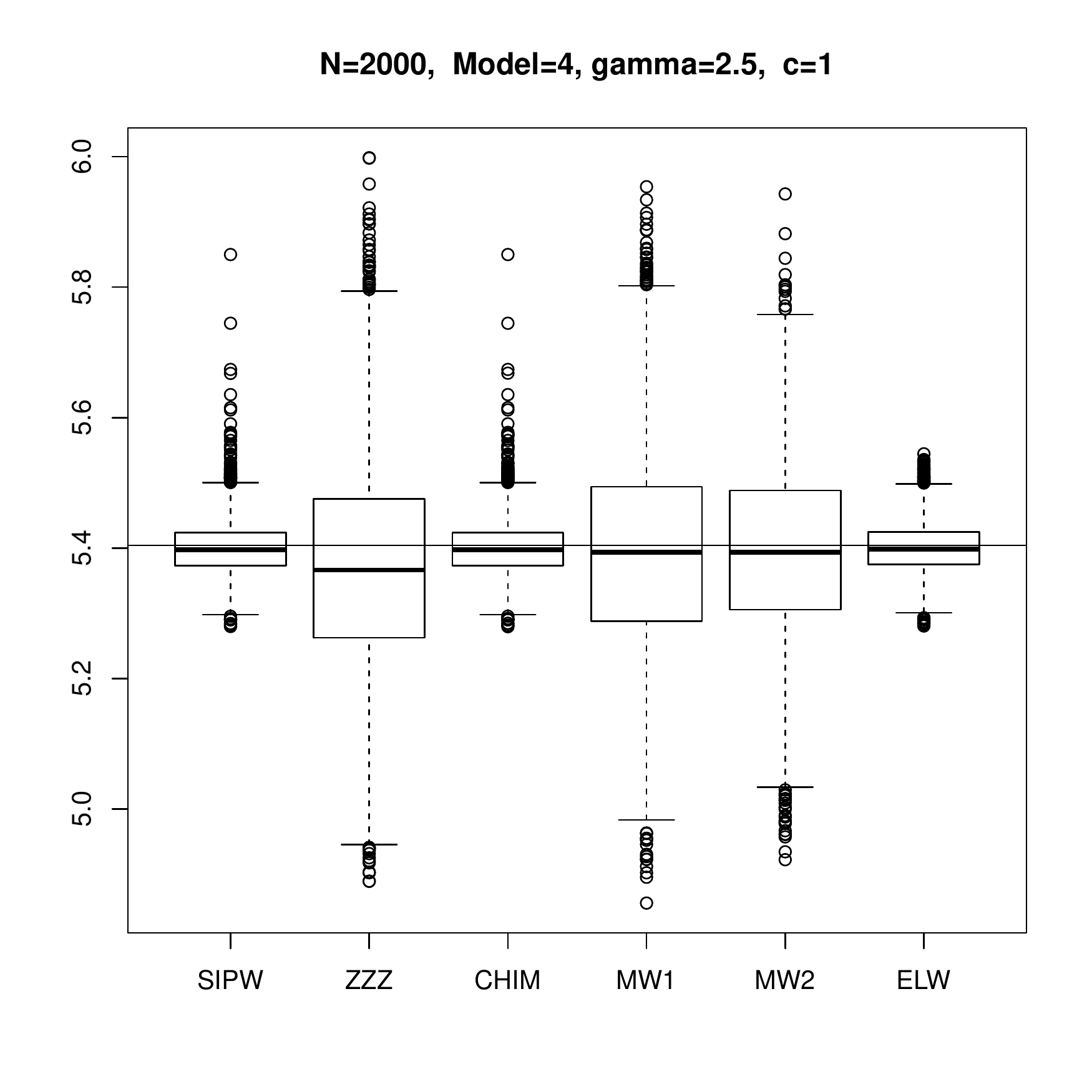}
\includegraphics[width=0.43\textwidth, height=0.23\textheight]{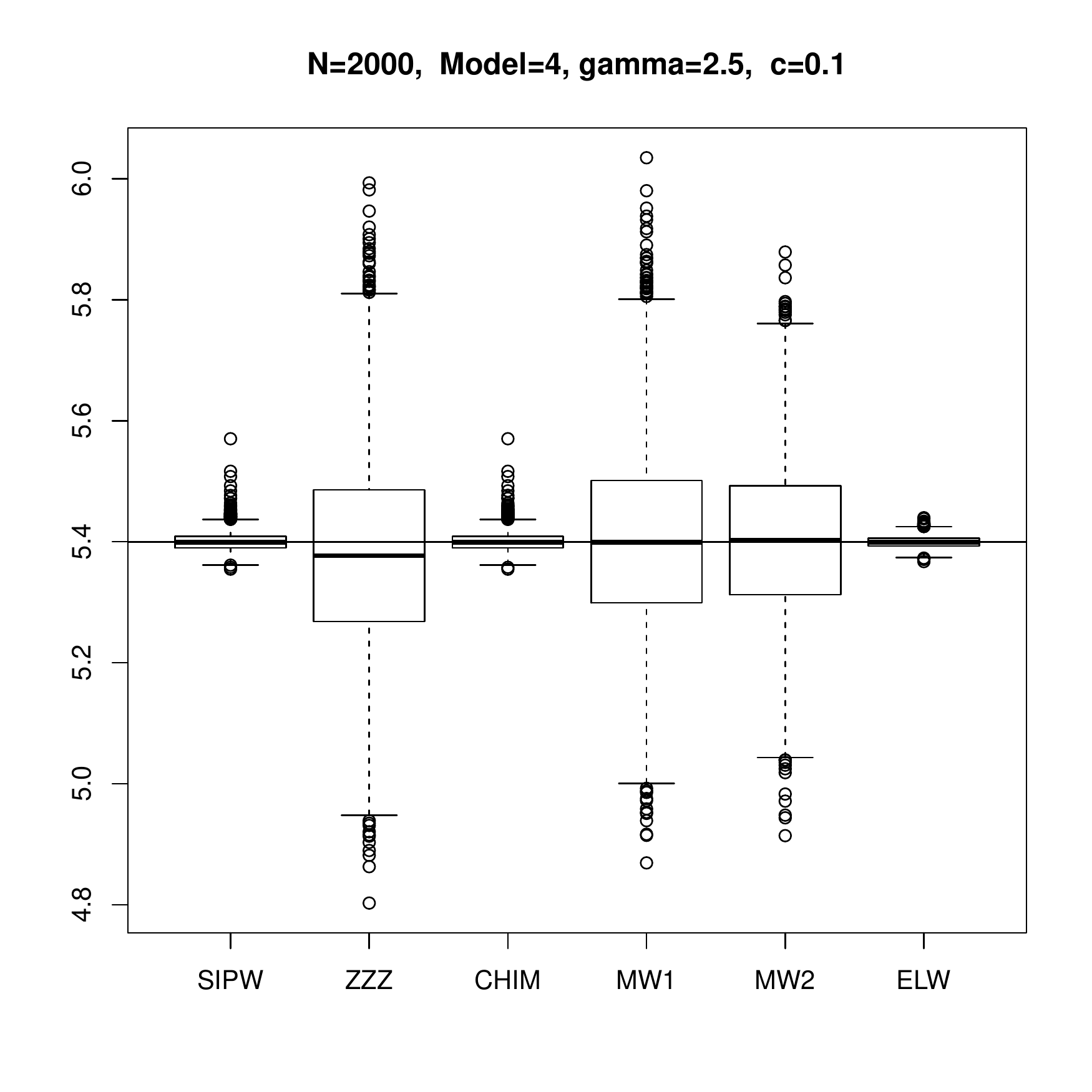} }
\caption{Boxplots of the SIPW, ZZZ, CHIM, MW1, MW2, and ELW estimators
when data were generated from  Example \ref{ex1} with  $N=2000$ and   $\gamma=2.5$.
The solid horizontal line corresponds to the target parameter value.
\label{fig-ex1-gamma=2.5} }
\end{figure}

{\it Point estimation }
As a measure of  the finite-sample performance of a generic estimator $\tilde \theta$,
we define its  scaled root mean square error (RMSE)   as
$
{\rm RMSE}(\tilde \theta) = \sqrt{N}\times \{ (1/M) \sum_{j=1}^M  (\tilde \theta_j - \theta)^2 \}^{1/2}
$,
where $\tilde \theta_j$ is the estimate
$\tilde \theta$ based on the $j$th simulated random sample.
Table   \ref{tab-rmse-ex1} presents a comparison of the  RMSEs of  the seven estimators,
and Figures \ref{fig-ex1-gamma=1.5} and \ref{fig-ex1-gamma=2.5}
 display the corresponding boxplots for $\gamma=1.5$ and 2.5, respectively.
For clearer presentation,   we ignore  the boxplots of  the IPW estimator,
because it fluctuates too dramatically.

In terms of RMSE, ELW   outperforms  IPW, SIPW, ZZZ, and CHIM
in almost all scenarios, which is clearly confirmed by their boxplots.
The only exception is   the scenario with $\gamma=2.5, c=1$ for Model  1,
where the RMSE (1.17) of ELW is slightly greater than the minimum RMSE (1.14)
of  IPW, SIPW, ZZZ, and CHIM.
The boxplots also indicate that  ELW  is always nearly unbiased
in all scenarios. ELW also  outperforms   MW1 and MW2 in most cases.
The only exceptions are the scenarios with $\gamma=2.5$ for Model 1
and those with $\gamma=1.5, c=1$ for Models 1 and  2.
In the least favourable scenario ($\gamma=1.5, c=1$, Model  2),
the RMSE of ELW is greater than those of MW1 and MW2 by at most $(5.13-3.78)/3.78 \approx 35.7\%$.
By contrast,  the RMSEs of  MW1 and MW2
can be more than 12 times  that of ELW;
see the scenario with $\gamma=1.5$ and $c=0.1$ for Model 4.
Although MW1 and MW2 often have smaller RMSEs than ELW for Model 1,
the boxplots in Figures \ref{fig-ex1-gamma=1.5} and \ref{fig-ex1-gamma=2.5}
indicate that they tend to have either non-ignorable biases or larger variances.

Models 3 and 4 are simply Models 1 and 2 with a mean shift.
When we change Models 1 and 2 to Models 3 and 4, respectively, and
keep the remaining settings unchanged, the boxplots demonstrate that
ELW clearly  performs the best:
it  not only is nearly unbiased, but also has the smallest variance.
Meanwhile,  ELW, CHIM, and SIPW have nearly unchanged RMSEs. This makes sense,
because their weights all sum to 1.
Unfortunately,  IPW, ZZZ, MW1, and MW2  are all very sensitive to a mean shift
in the data generating process, since their weights do not sum to 1.

When $c$ decreases from 1 to 0.1,  the influence of random error
become negligible and we expect all methods to exhibit  better performance.
Indeed, all methods  have decreasing RMSEs,   except for IPW.
ELW has the largest rates of decline in RMSE:
these rates are at least 69\% and 42\% when $\gamma=1.5 $ and $2.5$, respectively.
However, the RMSEs of ZZZ, MW1, and MW2 have nearly no reduction for Models 3 and 4.
ELW performs in the most stable manner, whereas the other methods have  either
extremely large fluctuations  or remarkable biases.

When $\gamma$ increases from $1.5$ to 2.5,
 ELW clearly outperforms the competitors in all scenarios except those for Model 1.
All methods exhibit similar  performance for Models 1 and 2.
However for Models 3 and 4,
IPW, ZZZ, MW1, and MW2  have much larger fluctuations,
compared with their performance for both Models 1 and 2.
This indicates that they are sensitive to a mean shift,
which is  undesirable.

Roughly speaking, among the seven estimators under comparison, the ELW estimator is the most reliable in almost all scenarios.
Both the RMSE results and the boxplots indicate that
MW1 and MW2 can exhibit very different performances.
In other words,  the performance of the method of \cite{Ma2020}
can be affected by the choice of  underlying tuning parameters.
We have also conducted simulations for $N=50$ and 500,
and $\gamma=1.3$ and 1.9. See Section 8 of the supplementary material
for the RMSE results and the corresponding boxplots.
The general findings are similar.

{\it Interval estimation }
Two confidence intervals for $\theta$ can be constructed based on  the ELW estimator $\hat \theta_{\elw}$.
One is the Wald confidence interval (ELW-an for short)
based on the asymptotic  normality of $\hat \theta_{\elw}$, where the asymptotic variance
is estimated using the ELW method.
The other is the resampling-based interval estimator (ELW-re) given in  Section \ref{sec2.5}.
Similar intervals (IPW-an, IPW-re, SIPW-an, and SIPW-re) can be constructed
when  the IPW and SIPW estimators take the place of the ELW estimator
in the estimations of both $\theta$ and the asymptotic variances.
We compare these six confidence intervals with those of \cite{Ma2020}
based on their resampling method and the MW1 and MW2 point estimators,
which are denoted by MW1-re and MW2-re, respectively.

We generate  random samples of size $N=2000$ from Example \ref{ex1},
and calculate the coverage probabilities and average lengths
of the eight confidence intervals at the 95\% confidence level.
The results are tabulated in Table \ref{cov-ex1}.
MW2-re has the most accurate coverage accuracy, followed by
 ELW-re  when $\gamma=1.5$ and
$c=1.0$ for Models 1 and 2.
When  Models 1 and 2 are replaced by  Models 3 and 4,
the coverage probabilities of ELW-re remain nearly unchanged; however,
those for MW1-re and MW2-re  decrease sharply by more than  5\% and 10\%, respectively.
When $c$ decreases from 1.0 to 0.1,
the coverage accuracy of ELW-re becomes better or is still acceptable,
although both MW1-re and MW2-re perform much more poorly.
With different tuning parameters, MW1-re and MW2-re
often have quite different coverage probabilities
and average lengths, which again shows that the performance of the method of
\cite{Ma2020}
can be greatly affected by different choices of tuning parameters.
SIPW-re has very close coverage probabilities
to  ELW-re in most cases, whereas its average lengths are generally
much greater than those of the latter.

As expected, all asymptotic-normality-based Wald intervals
exhibit remarkable undercoverage
when $\gamma=1.5$, because the asymptotic normalities
are  generally violated. In the meantime, all resampling-based intervals have improved performance.
When  $\gamma$ increases to 2.5,
all intervals except MW1-re and MW2-re  have very desirable coverage accuracy,
and   the asymptotic-normality-based intervals
have close or even better coverage probabilities compared with
the resampling-based intervals.

\begin{example}[Unequal probability sampling]
\label{ex2}
The finite population consists of $N=3000$ observations
$\{(x_i, y_i):  1\leq i\leq N\}$.
We generate  $x_i$  from the uniform distribution on  $[0, 2],$
and $y_i = \mu(x_i) +  \sqrt{3(1-\rho^2)} e_i,$
where   $e_i \sim N(0, 1)$ are independent of each other
and  of $x_i$.
We consider four models for $\mu(x)$, namely
  $ \sqrt{3} \rho x$  (Model 1),
  $ \sqrt{3} \rho (x+x^2)$  (Model 2),
  $ \sqrt{3} \rho x + 5$  (Model 3), and
  $ \sqrt{3} \rho (x+x^2)+5$  (Model 4),
and three unequal probability samplings, namely
Poisson sampling,   pivotal sampling,
and PPS (probability proportional to pize) sampling.
We fix $n=500,$  take $\pi_i = n x_i/\sum_{j=1}^N x_j$
for Poisson sampling and pivotal sampling,
and take $x_i$ as the size for  PPS sampling.
The goal is to estimate $\theta = (1/N) \sum_{i=1}^N y_i$.
\end{example}

Example \ref{ex2} is a modified version of Example 2 from \cite{Zong2019} and
is designed to represent unequal probability sampling.
Poisson sampling and pivotal sampling are UPS-WORs, with
 the sample size being random in the former and fixed in the latter.
PPS sampling is a popular UPS-WR.
In particular, pivotal sampling satisfies the NA  property
\citep{Branden2012} and
therefore  must satisfy the  LIND condition
\citep{Patterson2001}   required by Theorem \ref{CLT-B-ups}.

\begin{figure}
\centering
\mbox{
\includegraphics[width=0.43\textwidth, height=0.23\textheight]{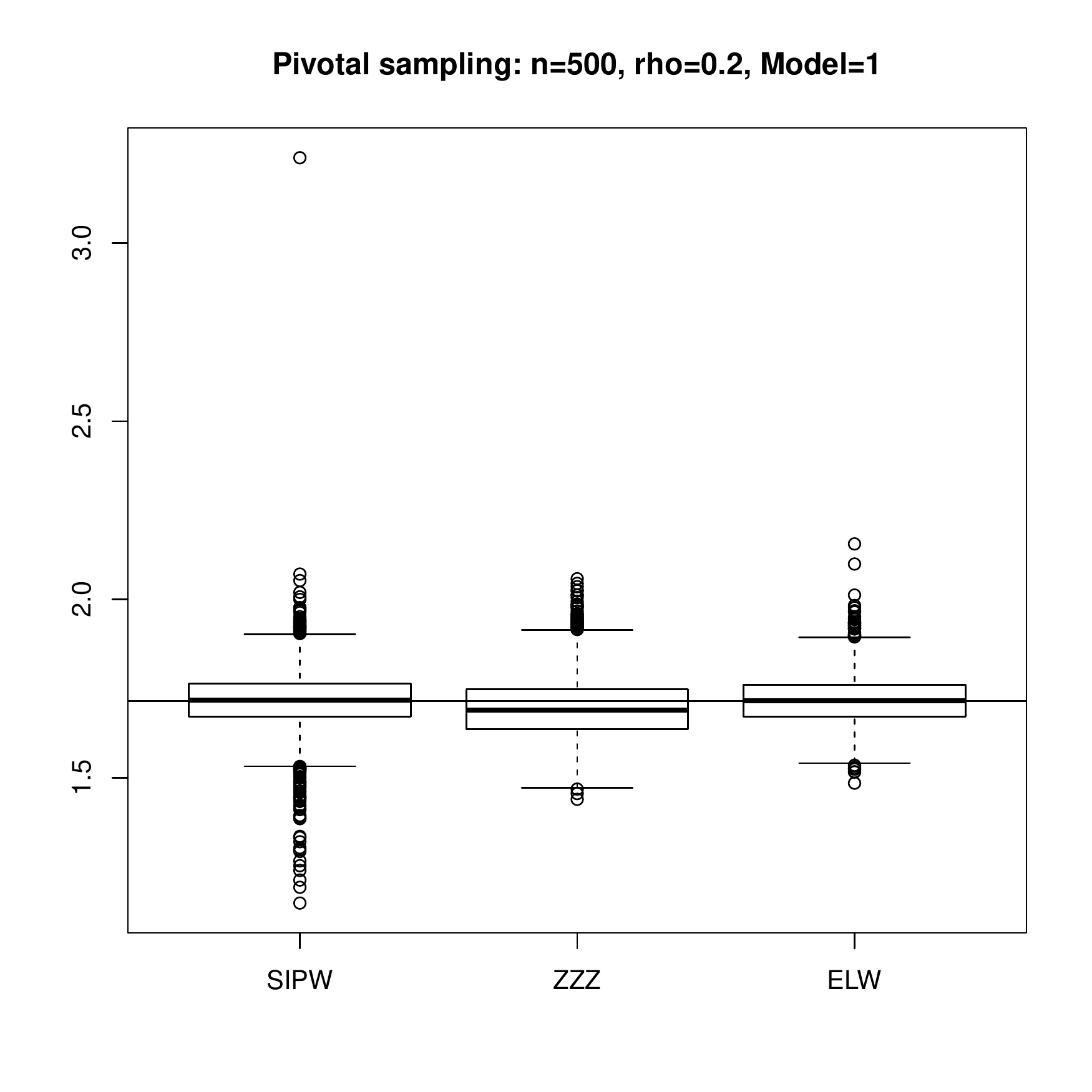}
\includegraphics[width=0.43\textwidth, height=0.23\textheight]{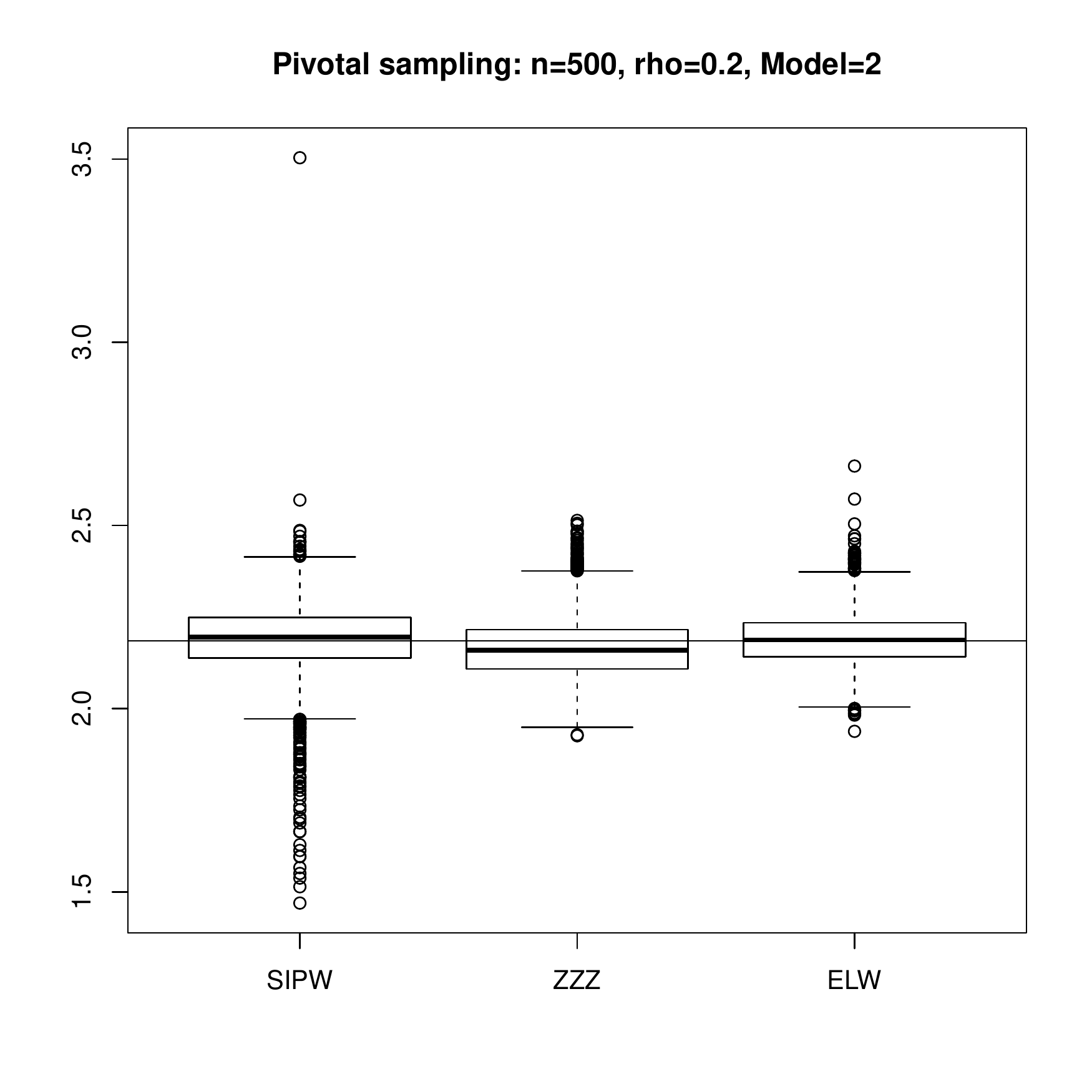} }  \\
\mbox{
\includegraphics[width=0.43\textwidth, height=0.23\textheight]{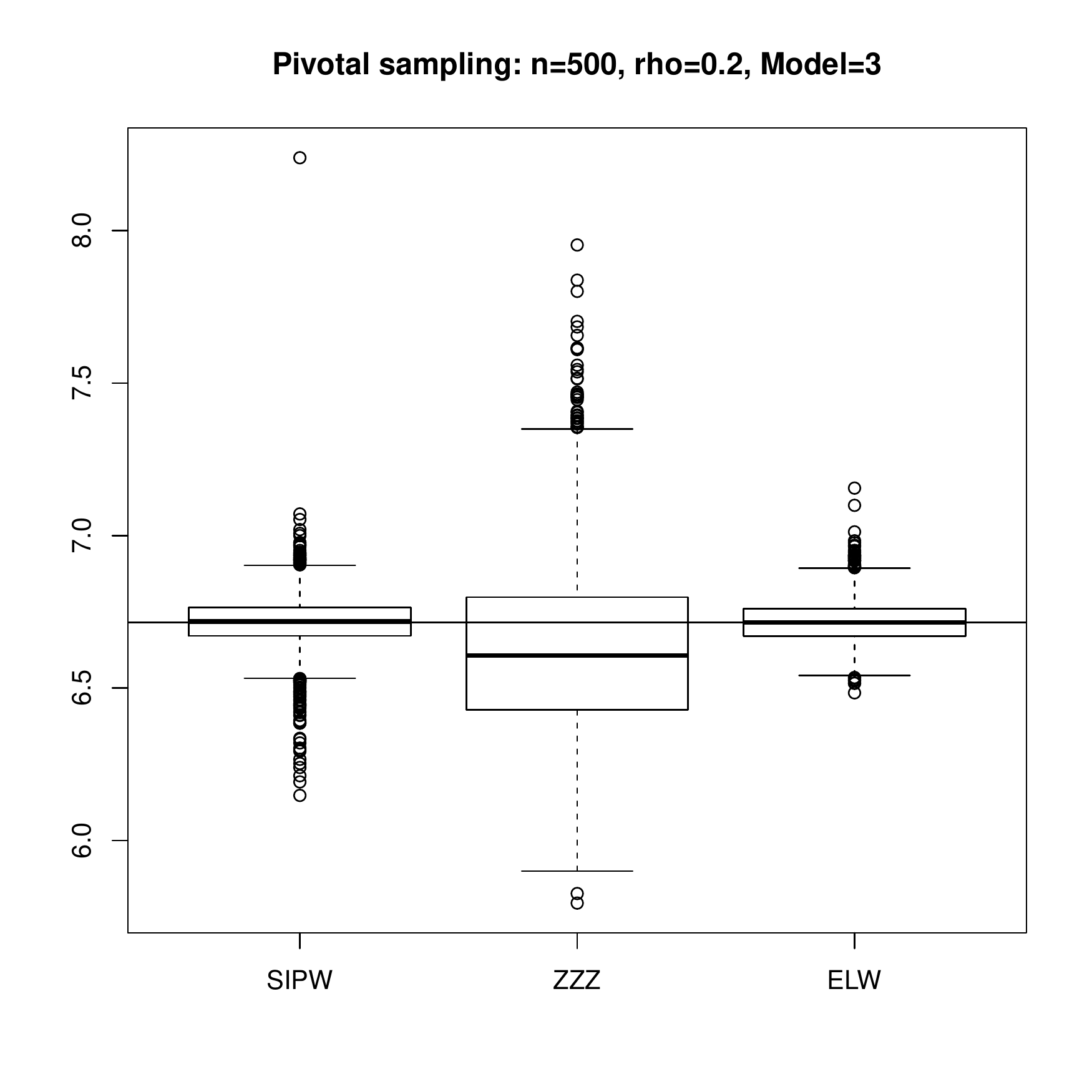}
\includegraphics[width=0.43\textwidth, height=0.23\textheight]{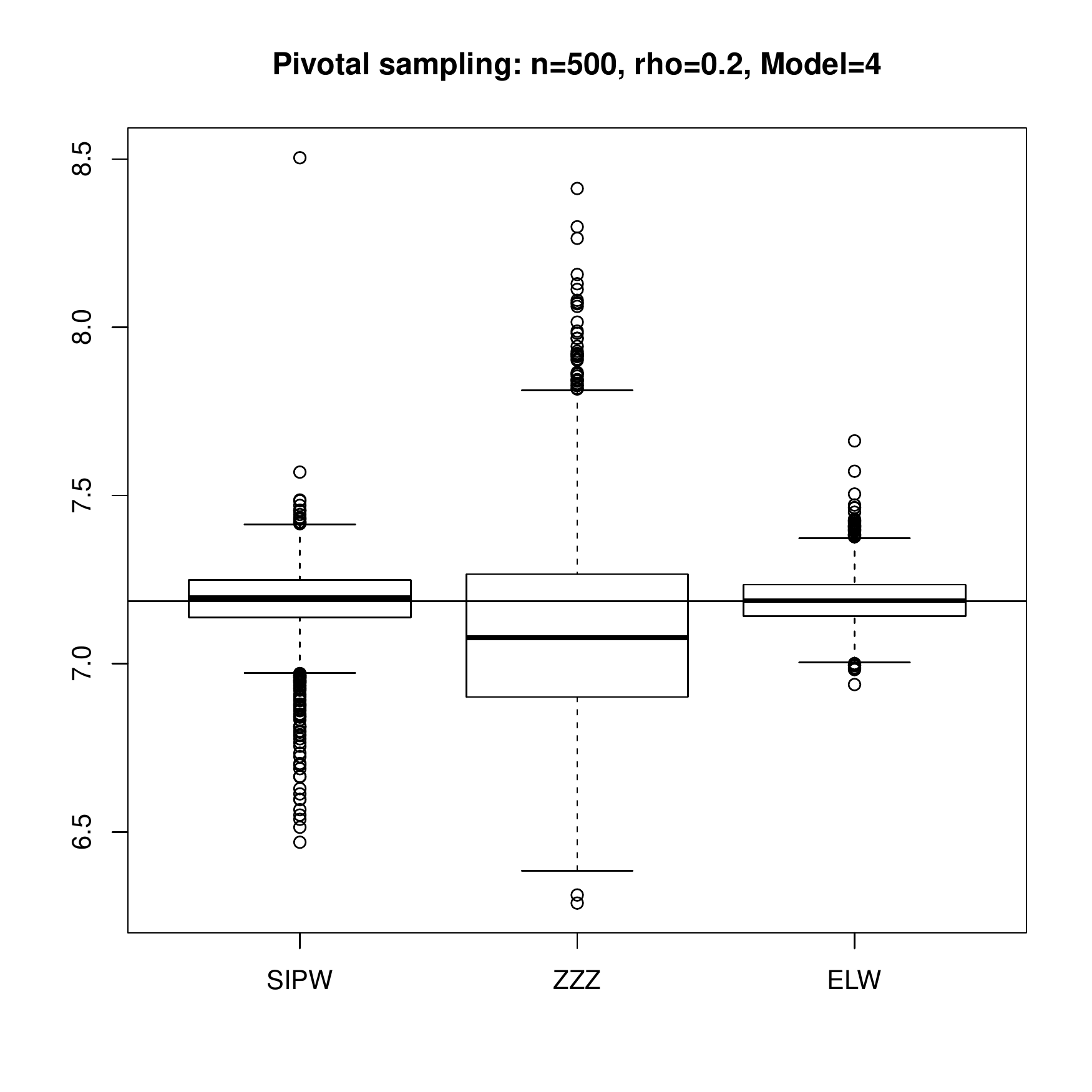} } \\
\mbox{
\includegraphics[width=0.43\textwidth, height=0.23\textheight]{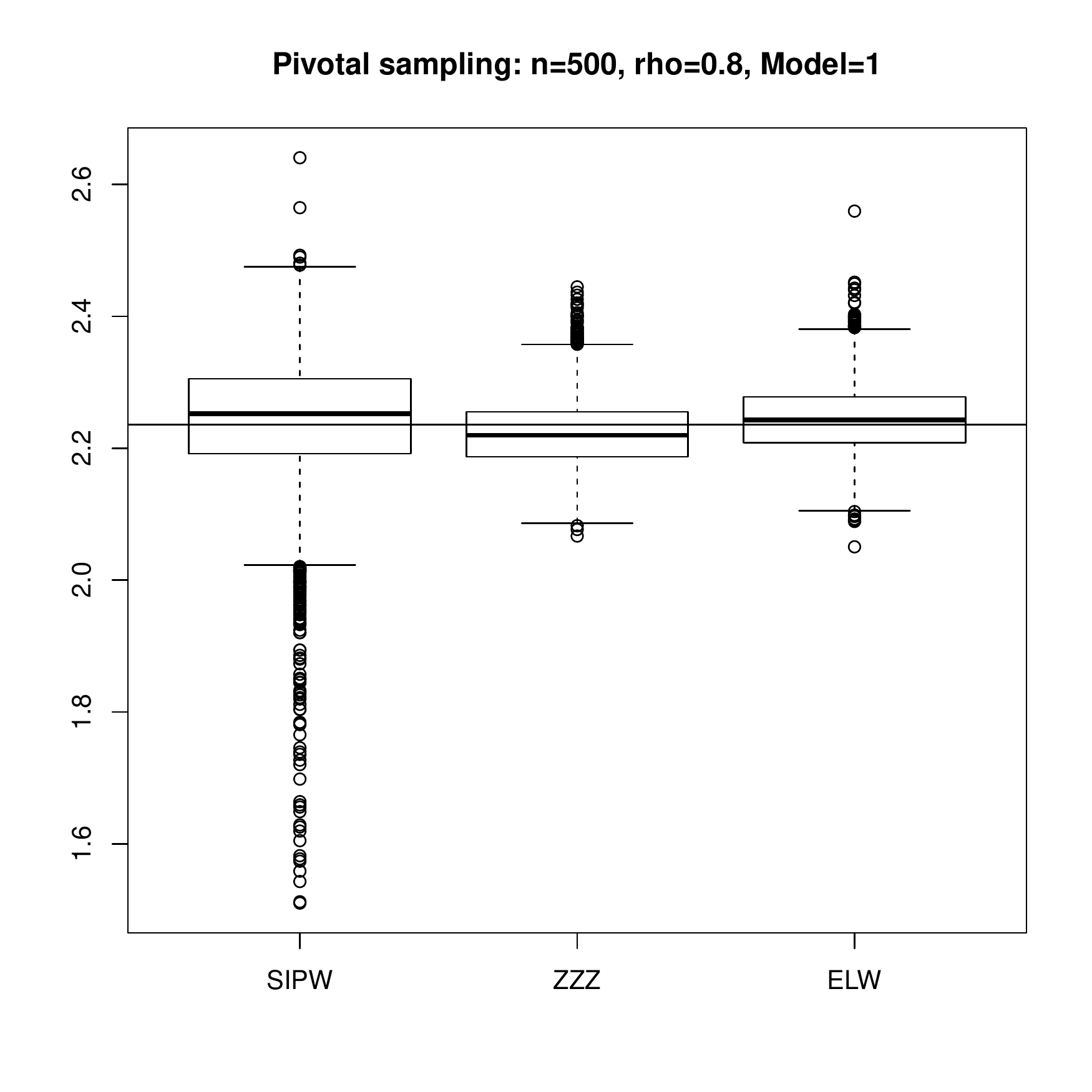}
\includegraphics[width=0.43\textwidth, height=0.23\textheight]{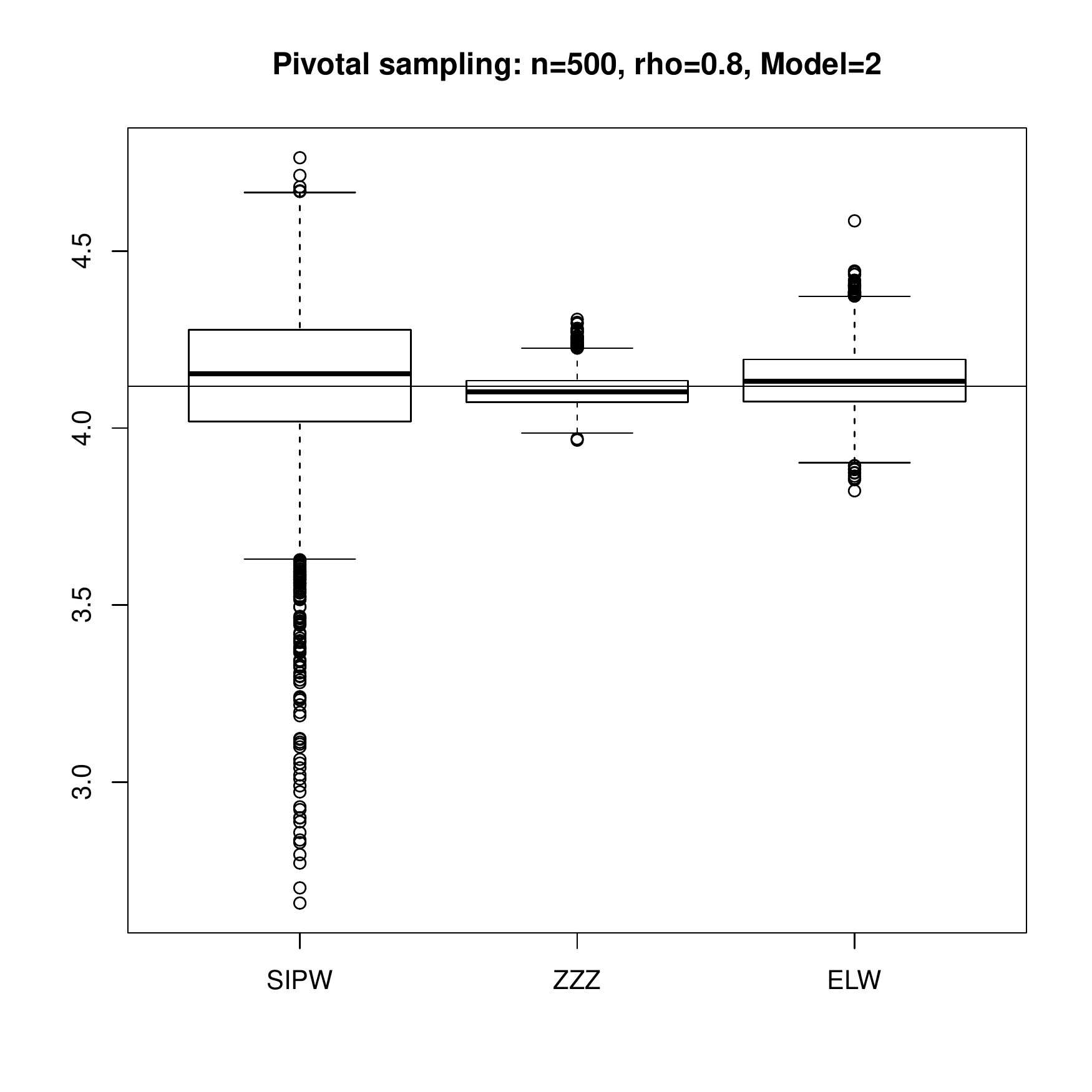} }  \\
\mbox{
\includegraphics[width=0.43\textwidth, height=0.23\textheight]{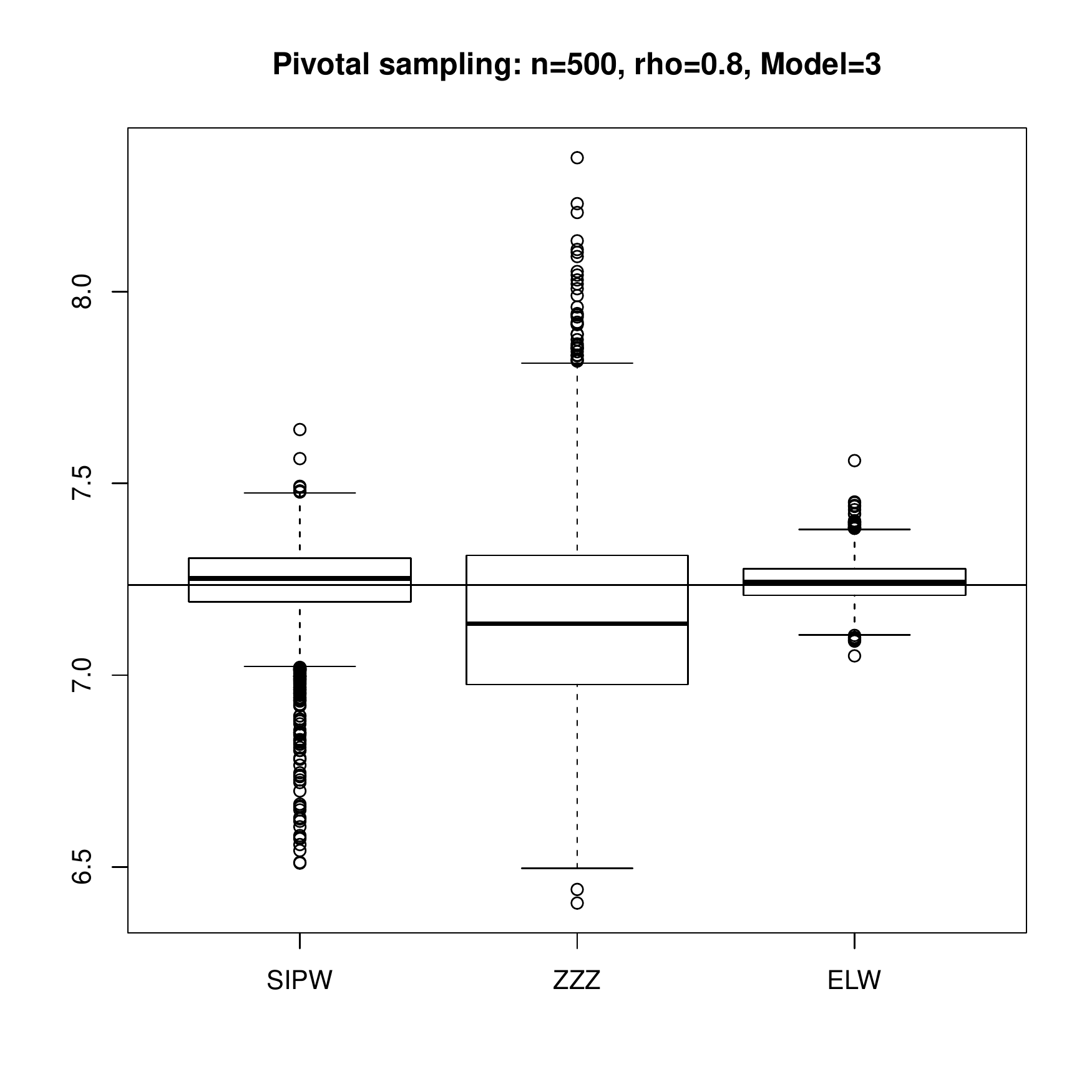}
\includegraphics[width=0.43\textwidth, height=0.23\textheight]{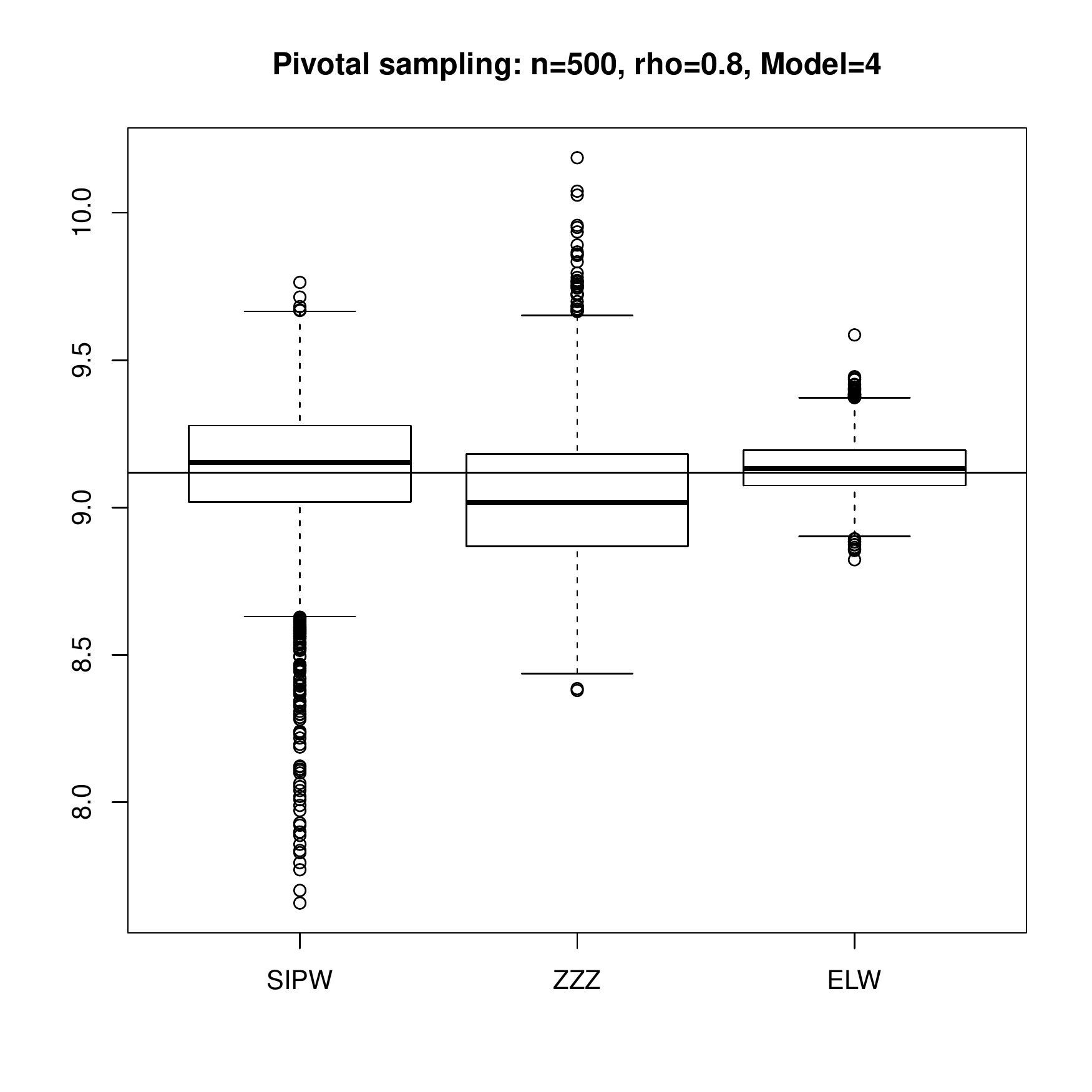} }
\caption{
Boxplots of the SIPW, ZZZ,  and ELW estimators
when data were generated by pivotal sampling  from  Example \ref{ex2} with $n=500$.
The solid horizontal line corresponds to the target parameter value.
\label{ex2-Pitoval-500}
}
\end{figure}

We exclude CHIM, MW1, and MW2 from the subsequent comparison
because they are all designed not for  unequal probability sampling
but for missing data problems.
Table \ref{tab-rmse-ex2-500} presents the simulated RMSEs for
the IPW, SIPW, ZZZ, and ELW estimators,
and Figure \ref{ex2-Pitoval-500}  displays the accompanying boxplots
under pivotal sampling.
The boxplots under  Poisson and PPS samplings are similar,
and can be found in the supplementary material.

ELW always has the smallest RMSEs under Poisson sampling.
It still perform the best under both pivotal and PPS samplings,
except  for the scenarios with $\rho=0.8$ for Models 1 and 2.
When the RMSE of ELW is not the smallest,
the boxplots of ZZZ and ELW are close to each other.
In the remaining scenarios, the competitors of ELW have  either much larger variances
or much larger biases than ELW.
Similar to Example \ref{ex1},  Models 3 and 4 in Example \ref{ex2} are
also Models 1 and 2 with a mean shift.
Again, the performance of ELW and SIPW is equivariant under such a mean shift;
however, the IPW and ZZZ estimators are very sensitive, and their performances
become much worse when the mean  of $y_i$ moves away from the origin.
The performances of the four estimators seem to be insensitive to the choice of $\rho$.

For interval estimation under unequal probability sampling,
no valid resampling-based confidence intervals are available as far as we know.
Therefore, we compare only IPW-an, SIPW-a, and ELW-an,
which are the  confidence intervals
based on the asymptotic normalities of the IPW, SIPW, and ELW estimators.
Table \ref{tab-cov-ex2-500} presents their simulated coverage probabilities
and average lengths.
Under both Poisson and pivotal samplings,  overall
ELW-an has not only  more accurate  coverage probabilities
but also  shorter average lengths than SIPW-an.
Under PPS sampling, when $\rho=0.8$,
ELW-an still wins, although both ELW-an and SIPW-an have undercoverages of 3\% or more.
However, when $\rho=0.2$, SIPW-an has better and more desirable
coverage probabilities. This implies that ELW-an still has room for improvement.

In summary, the ELW point estimator  has the most reliable overall
performance,  is shift-equivariant, and is nearly unbiased in all cases.
The proposed resampling-based ELW interval estimator often has desirable
coverage accuracy and short lengths in  missing data problems,
whether the proportion of extremely small propensity scores is small or large.
The asymptotic-normality-based ELW interval estimator
has desirable performance under UPW-WORs such as Poisson and pivotal samplings,
and acceptable performance under UPS-WRs such as PPS sampling.

 \section{Real data analysis}\label{sec5}

\cite{Lalonde1986} estimated the impact of the National Supported Work Demonstration, a labour
training programme, on post-intervention income levels, using data from a randomized evaluation
of the programme. To further demonstrate the superiority of the proposed ELW method,
we analyse the  {\tt LLvsPSID} data from the {\tt R} package {\tt cem},
which is the Lalonde set of treated units versus PSID
(Panel Study of Income Dynamics) control individuals.
The data consist of 2787 observations (297 from treated units and 2490 from control units)
on 10 variables:  {\it treated} (treatment indicator),
{\it age} (age),
{\it education} (years of education),
{\it black} (race, indicator variable),
{\it married} (marital status, indicator variable),
{\it nodegree} (indicator variable of not possessing a degree),
{\it re74} (real earnings in 1974),
{\it re75} (real earnings in 1975),
{\it re78} (real earnings in 1978),
{\it hispanic} (ethnic, indicator variable),
{\it u74} (unemployment in 1974, indicator variable), and
{\it u75} (unemployment in 1975, indicator variable).
The variable {\it re78}  is the  post-treatment outcome.

Let $Y = ${\it re78}/10\,000 be the response, let $D=$ {\it treated},
and let $Y(d)$
denote the response of an individual whose treatment status is $D=d$.
We shall not  address the original
treatment effect estimation problem. Instead, we
take the data as  missing data
and wish to estimate the average earnings
of the treated in 1978.
In other words, the parameter of interest is   $\theta=\e\{ Y(1)\}$.
We first estimate  the propensity scores by fitting
a linear logistic regression model of the treatment indicator $D$
on the remaining eight variables (excluding $D$, $Y$, and {\it re78}).
With the fitted propensity scores,
the IPW,  SIPW,  MW1,  MW2, and   ELW point estimates are
0.65, 0.92, 0.72, 0.70, and 1.11, respectively, and
the corresponding resampling-based interval estimates at the 95\% level
are $[-12.68,  8.00]$,
$[-4.16,  3.15]$,
$[-3.04,  1.26]$,
$[-0.27, 1.04]$, and $ [-8.86,  6.27]$, respectively.
If we  replace  all $Y$   by $Y+5$,
the  point estimates become
4.16, 5.92, 5.81, 5.56, and 6.11
with   interval estimates being   $[-27.31,  22.76]$,
$[0.90,  8.16]$,
$[0.92, 7.16]$,
$[5.05, 6.22]$, and $ [-3.87,  11.44]$, respectively.
As expected, the SIPW and ELW point estimates are shift-equivariant, but
  the IPW estimator and  the MW estimators are not.

\begin{figure}
\centering
\mbox{
\includegraphics[width=0.3\textwidth ]{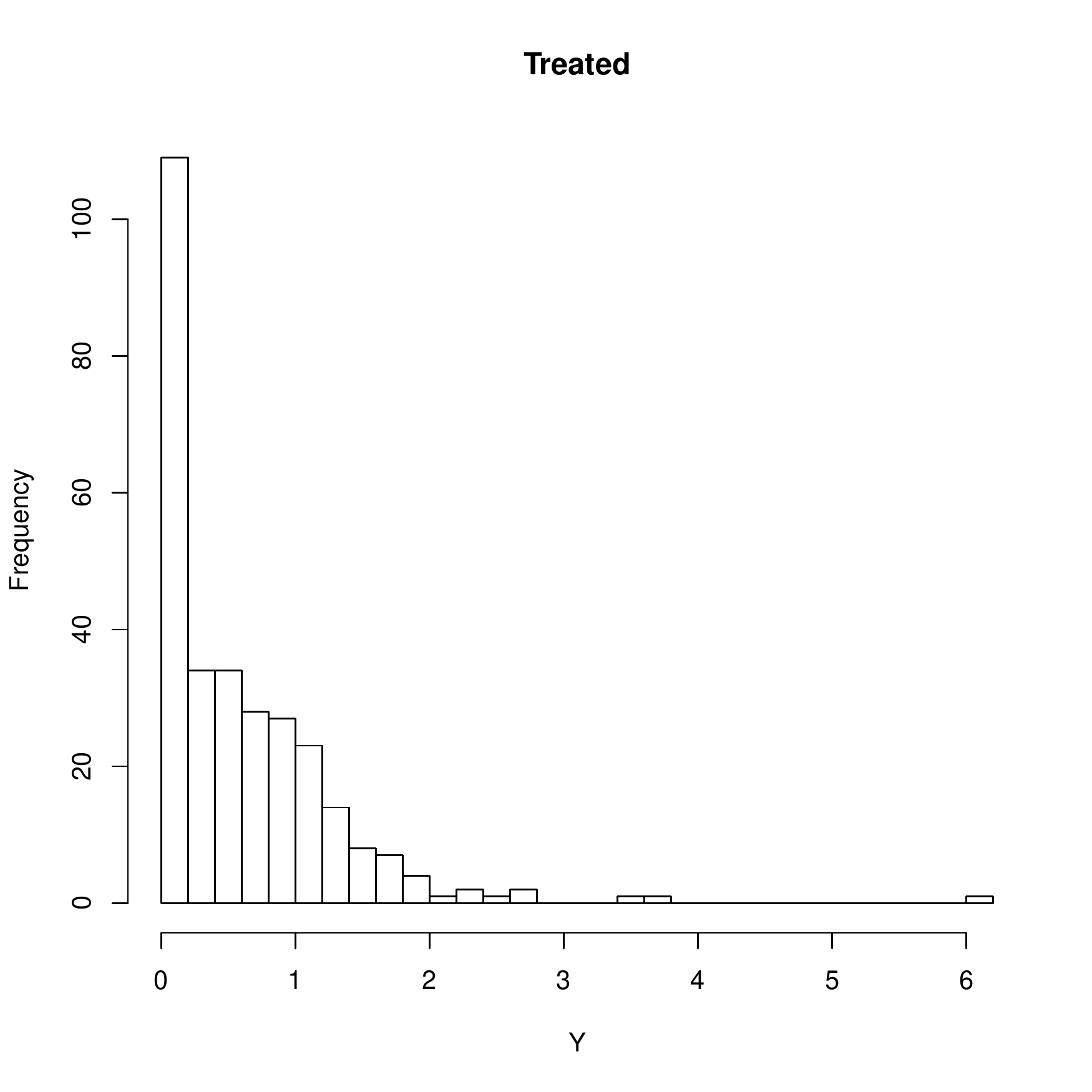}
\includegraphics[width=0.3\textwidth ]{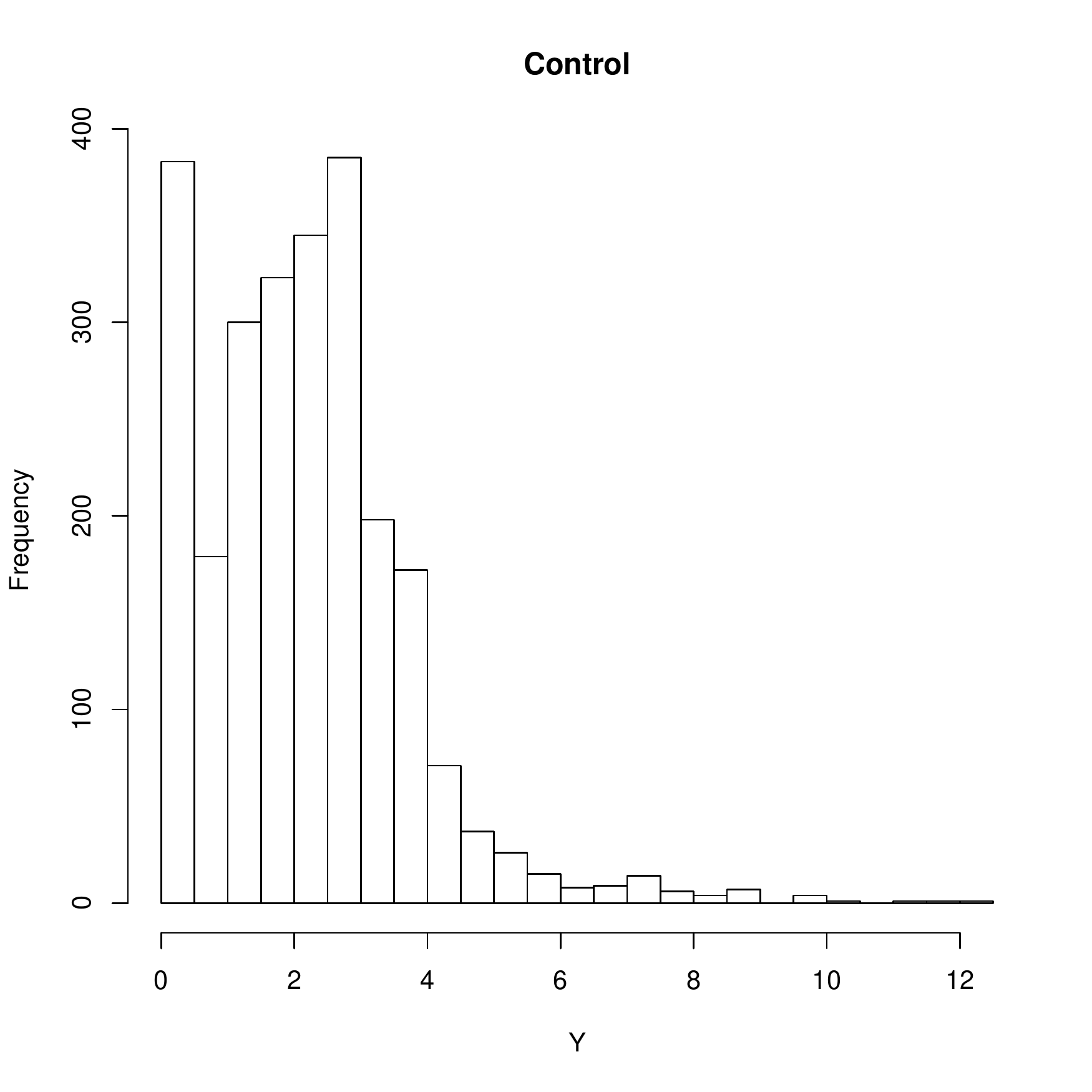}} \\
\mbox{
\includegraphics[width=0.3\textwidth ]{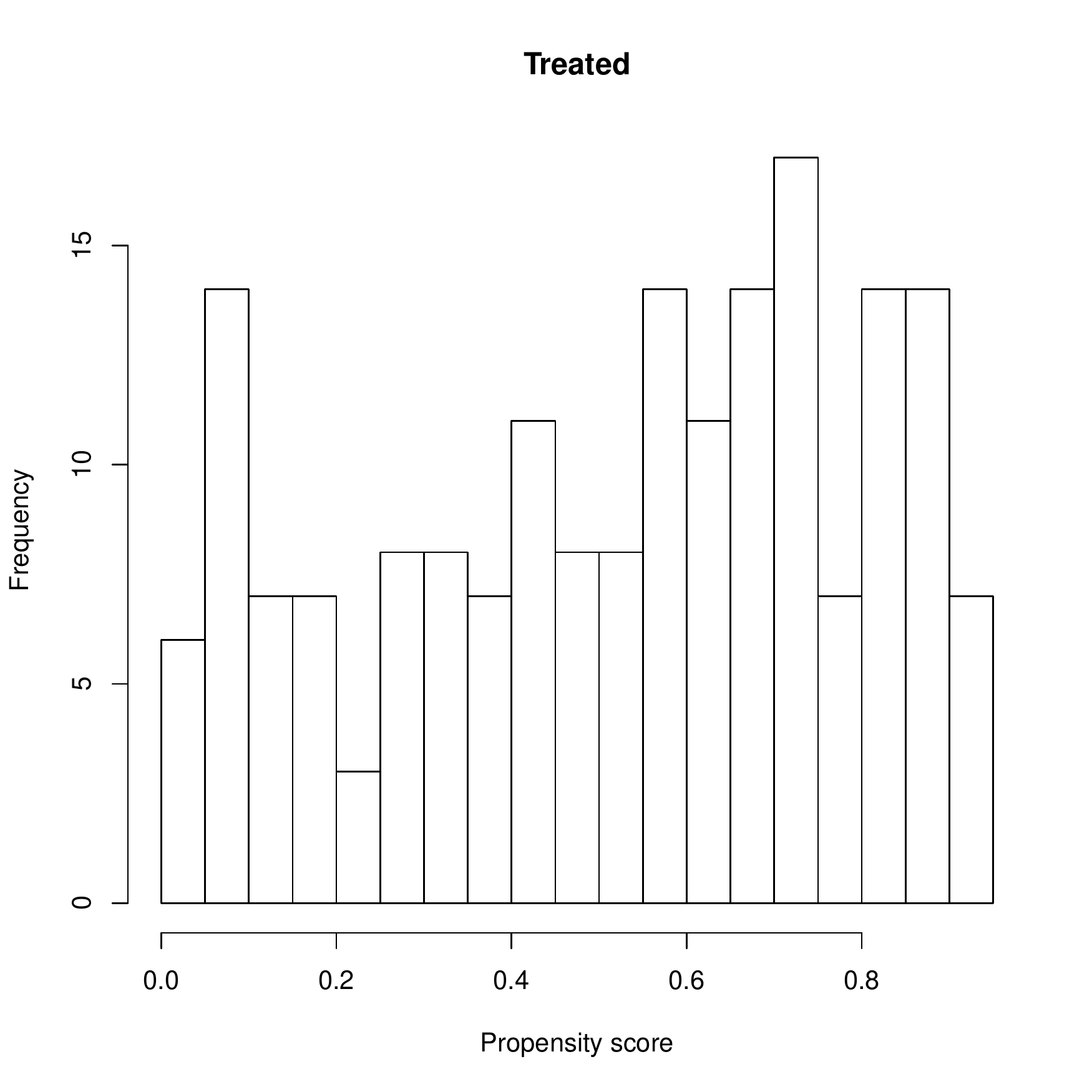}
\includegraphics[width=0.3\textwidth ]{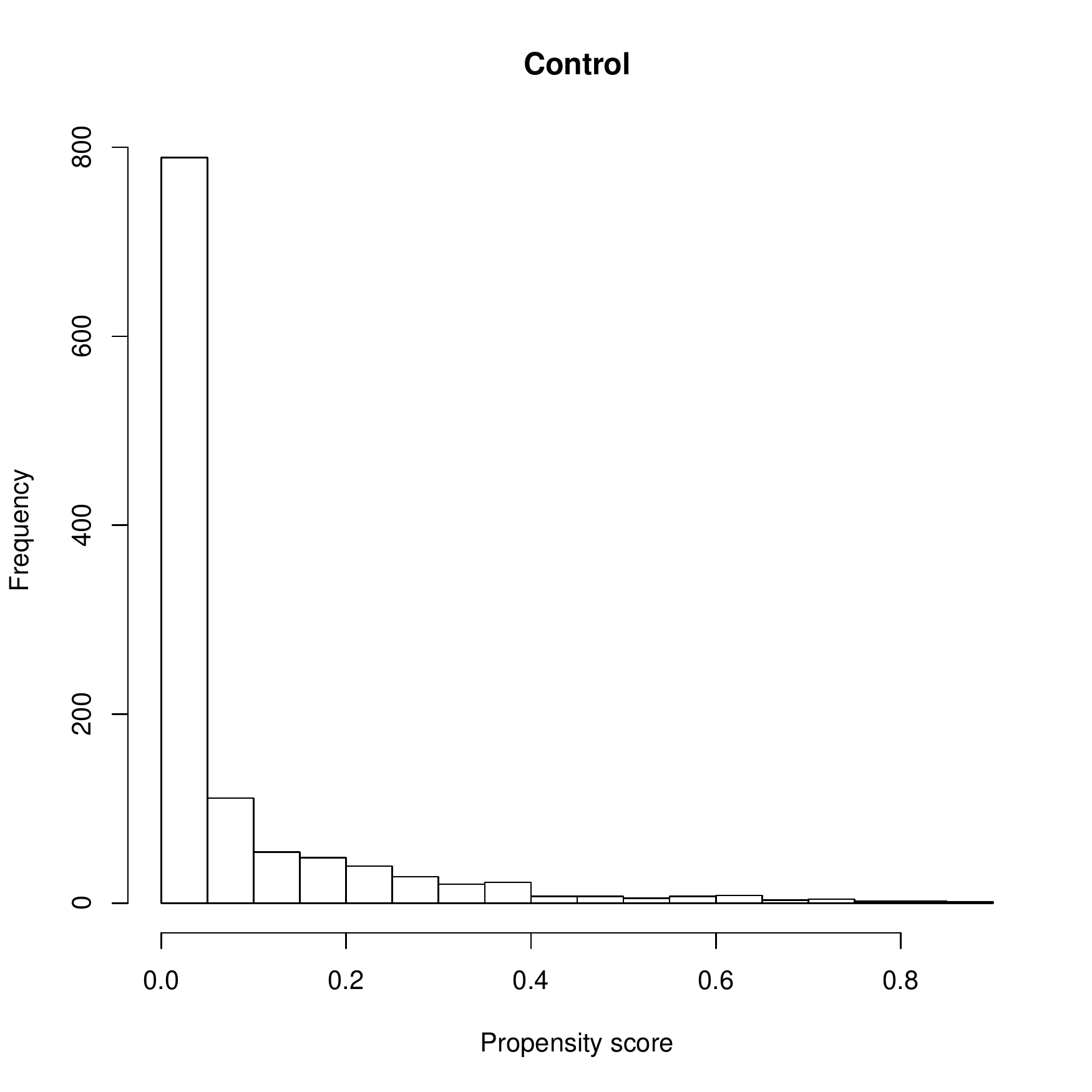}}
\caption{
\label{yps-realdata}
 Histograms of the variable  $Y$  and
the fitted propensity score  in
the treated and control groups,
based on the {\tt LLvsPSID} data.  }
\end{figure}

Figure \ref{yps-realdata} displays the fitted propensity scores
of both  the treated and control groups.
A clump of near-zero  propensity scores
in the treated group implies that the standard
IPW estimator is dramatically unstable.
The  excessive number of near-zero  propensity scores  in both groups
indicates that the distribution of the inverse propensity score
has a very heavy right tail  similar to that
in the simulation scenario with $\gamma=1.5$ in Example \ref{ex1}.
According to our simulation experience in the case of $\gamma=1.5$,
the ELW point estimator is always unbiased or nearly unbiased,
and its performance is the most stable in most cases.
By contrast, the other estimators  SIPW, MW1, and MW2
may have either much larger  RMSEs or large biases.
The ELW-re interval has the most desirable and much better
coverage accuracy than the other intervals.
These observations makes it reasonable to believe that
the ELW point and interval estimates, 1.11  and $ [-8.86,  6.27]$,
are  the most preferable for the estimation of $\theta = \e\{ Y(1)\}$.

In addition, we regard the observations in  the  {\tt LLvsPSID} data
with non-zero {\it re75}  as a finite population,
and conduct Poisson, pivotal, and PPS samplings
with inclusion probabilities proportional to {\it re75}.
We  take the parameter of interest to be the mean of $Y = {\it re78}/10\,000 + a$,
with $a=0$ or 2.
Table \ref{tab-rmse-realdata} presents the simulated RMSE results
based on 5000 simulation repetitions
with a sample size   (in pivotal and PPS samplings) or ideal sample size
(in Poisson sampling) of $200$.
The ELW estimator  has the smallest RMSEs under Poisson sampling, regardless of whether $a=0$ or 2,
and  under  pivotal and PPS samplings when $a=2$.
It also uniformly outperforms SIPW under all three samplings.
When $a=0$,  its performance can be inferior to those of IPW and ZZZ, which,
 however, are highly sensitive to a location  shift in $Y$.
The ELW estimator  again   has
the best overall performance under unequal probability sampling.

\section{Discussion}\label{sec6}

When introducing the ELW method,
we assume that the parameter of interest
has a simple form  $\theta = \e\{g(Z)\}$,
for convenience of presentation.
Our method can be directly extended to the parameters defined
by a group of estimation equations \citep{Qin1994}.
Much more effort needs to be applied to
 the investigation of the large-sample properties of the
corresponding ELW estimator.

In the development of the  ELW method,
we regard $\alpha$ as unknown
to guarantee that the EL weights are always well defined.
In some situations, such as  Poisson  and PPS samplings,
the true parameter value $\alpha_0$ of $\alpha$ may be known a priori.
Under PPS sampling, the coverage accuracy of the ELW-based Wald interval is not quite satisfactory.
Hopefully, replacing $\alpha$ with $\alpha_0$ or $n/N$ may help,
although the EL weights may be undefined.
This poses the interesting problem of how to
overcome the non-definition problem of the EL method
while simultaneously improving its estimation efficiency.

When constructing confidence regions for the parameter $\theta$,
we consider only the Wald method in this paper.
In missing data problems or under Poisson sampling,
an alternative is to construct confidence intervals
based on the empirical likelihood ratio function, which necessitates
 study of the limiting distribution of the empirical likelihood ratio
statistic.
We believe that
this statistic asymptotically
follows a central chi-square distribution,
making it convenient for the construction of
empirical likelihood ratio regions with asymptotically correct
confidence level.

\section*{Supplementary material}
The supplementary material contains proofs of
Proposition \ref{prop-kappa},
Theorems  \ref{CLT-A-known},
\ref{CLT-A-unknown}, \ref{CLT-B-ups},
\ref{CLT-B-ups2}, and
\ref{CLT-C-ups}, and more simulation results.

\section*{ Acknowledgements}

The research  was supported by the National Natural Science Foundation
of China (12171157 and 11971300), the Natural Science Foundation of Shanghai  (19ZR1420900),
the State Key Program of the National Natural Science Foundation of China (71931004 and 32030063),
the Development Fund for Shanghai Talents, and  the 111 Project (B14019).

\begin{table}
 \caption{
 Simulated RMSEs of the estimators under comparison
 when data are generated from Example \ref{ex1} and $N=2000$.
 Smallest RMSEs are highlighted in bold.
  \label{tab-rmse-ex1} }
\centering
  \begin{tabular}{ c c c|rrrrrrr}
  \hline
   $\gamma$  &   $c$ &   Model    &     IPW &   SIPW & ZZZ    & CHIM  &   MW1 &   MW2 &  ELW \\  \hline

1.5& 1.0&     1&  24.72&  8.05&  6.05&  8.00&  4.96&  {\bf 4.35} &  5.51 \\
1.5& 1.0&     2&  17.89&  6.17&  5.95&  6.17&  4.84&  {\bf 3.78} &  5.13 \\
1.5& 1.0&     3&  69.08&  7.49& 27.27&  7.49& 18.29&  9.59&  {\bf 5.21} \\
1.5& 1.0&     4& 110.80&  6.49& 27.11&  6.49& 18.31&  9.83&  {\bf 5.21} \\
1.5& 0.1&     1&  14.76&  4.89&  4.48&  4.87&  3.00&  2.52&  {\bf 1.60} \\
1.5& 0.1&     2&  26.23&  2.16&  4.44&  2.15&  2.88&  1.36&  {\bf 0.71} \\
1.5& 0.1&     3&  68.12&  4.74& 27.04&  4.73& 17.81&  8.94&  {\bf 1.61} \\
1.5& 0.1&     4& 140.05&  2.21& 26.86&  2.19& 18.02&  9.03&  {\bf 0.74} \\   \hline
2.5& 1.0&     1&   2.11&  2.11&  1.97&  2.11&  1.93&  {\bf 1.87}&  2.02 \\
2.5& 1.0&     2&   2.06&  1.81&  1.90&  1.81&  1.89&  1.82 & {\bf  1.72} \\
2.5& 1.0&     3&   7.64&  2.15&  6.77&  2.15&  6.33&  5.32&  {\bf 2.05} \\
2.5& 1.0&     4&   8.14&  1.85&  7.31&  1.85&  6.87&  6.01&  {\bf 1.70} \\
2.5& 0.1&     1&   1.49&  1.33&  1.14&  1.33&  1.07&  {\bf 0.98}&   1.17  \\
2.5& 0.1&     2&   1.22&  0.69&  1.01&  0.69&  0.93&  0.77&  {\bf 0.42} \\
2.5& 0.1&     3&   7.63&  1.31&  6.60&  1.31&  6.21&  5.14&  {\bf 1.18} \\
2.5& 0.1&     4&   8.26&  0.68&  7.13&  0.68&  6.80&  5.85&  {\bf 0.42} \\
  \hline
  \end{tabular}
\end{table}

\begin{table}
 \caption{
Simulated coverage probabilities (\%)
of the interval estimators under comparison
when data were generated from Example \ref{ex1} with
total sample size $N=2000$.
 The numbers in parentheses are average lengths.
  \label{cov-ex1} }
 \tabcolsep 3pt
\centering
  \begin{tabular}{c cc|cc|cc|cc|cc }
  \hline
      $\gamma$   &   $c$  & Model    &  IPW-an   &   IPW-re   & SIPW-an   &   SIPW-re  &    MW1-re &   MW2-re     &   ELW-an   & ELW-re\\
     \hline
1.5& 1.0 &    1 & $\underset{(0.556)}{76.14}$&  $\underset{(   3.029)}{84.52}$&   $\underset{(0.436)}{78.42}$&   $\underset{(1.455)}{88.86}$& $\underset{(1.060)}{89.44}$& $\underset{(0.780)}{92.86}$&  $\underset{(0.329)}{82.58}$&  $\underset{(1.044)}{91.48}$  \\
1.5& 1.0 &    2 & $\underset{(1.593)}{78.24}$&  $\underset{(6397.733)}{87.24}$&   $\underset{(0.366)}{85.08}$&   $\underset{(1.159)}{91.14}$& $\underset{(1.165)}{89.68}$& $\underset{(0.831)}{94.52}$&  $\underset{(0.312)}{81.62}$&  $\underset{(1.047)}{91.04}$  \\
1.5& 1.0 &    3 & $\underset{(2.948)}{76.76}$&  $\underset{(  18.447)}{86.40}$&   $\underset{(0.446)}{77.84}$&   $\underset{(1.575)}{88.94}$& $\underset{(3.325)}{82.74}$& $\underset{(1.507)}{81.44}$&  $\underset{(0.333)}{82.02}$&  $\underset{(1.061)}{90.94}$  \\
1.5& 1.0 &    4 & $\underset{(3.282)}{78.74}$&  $\underset{( 121.859)}{86.32}$&   $\underset{(0.370)}{85.36}$&   $\underset{(1.160)}{91.60}$& $\underset{(3.276)}{82.32}$& $\underset{(1.551)}{83.32}$&  $\underset{(0.312)}{82.54}$&  $\underset{(1.046)}{92.38}$  \\
\hline
1.5& 0.1 &    1 & $\underset{(0.491)}{76.94}$&  $\underset{(   6.361)}{88.86}$&   $\underset{(0.304)}{79.58}$&   $\underset{(1.031)}{91.54}$& $\underset{(0.543)}{79.76}$& $\underset{(0.261)}{68.98}$&  $\underset{(0.126)}{92.86}$&  $\underset{(0.287)}{95.72}$  \\
1.5& 0.1 &    2 & $\underset{(0.477)}{76.22}$&  $\underset{(   5.052)}{87.68}$&   $\underset{(0.130)}{76.54}$&   $\underset{(0.436)}{89.20}$& $\underset{(0.552)}{81.92}$& $\underset{(0.224)}{82.40}$&  $\underset{(0.049)}{87.62}$&  $\underset{(0.122)}{91.72}$  \\
1.5& 0.1 &    3 & $\underset{(2.952)}{77.40}$&  $\underset{(  22.300)}{86.18}$&   $\underset{(0.304)}{78.00}$&   $\underset{(1.036)}{90.44}$& $\underset{(3.177)}{80.50}$& $\underset{(1.293)}{78.90}$&  $\underset{(0.125)}{91.84}$&  $\underset{(0.287)}{94.84}$  \\
1.5& 0.1 &    4 & $\underset{(2.958)}{77.88}$&  $\underset{(  31.359)}{85.96}$&   $\underset{(0.131)}{76.22}$&   $\underset{(0.441)}{89.00}$& $\underset{(3.125)}{81.76}$& $\underset{(1.361)}{79.12}$&  $\underset{(0.049)}{85.38}$&  $\underset{(0.122)}{89.28}$  \\
\hline
2.5& 1.0 &    1 & $\underset{(0.178)}{94.02}$&  $\underset{(   0.341)}{91.40}$&   $\underset{(0.177)}{93.60}$&   $\underset{(0.351)}{92.46}$& $\underset{(0.323)}{91.92}$& $\underset{(0.323)}{93.14}$&  $\underset{(0.169)}{93.32}$&  $\underset{(0.343)}{93.20}$  \\
2.5& 1.0 &    2 & $\underset{(0.169)}{93.52}$&  $\underset{(   0.346)}{93.48}$&   $\underset{(0.150)}{93.72}$&   $\underset{(0.303)}{93.54}$& $\underset{(0.324)}{92.38}$& $\underset{(0.321)}{93.40}$&  $\underset{(0.141)}{93.72}$&  $\underset{(0.287)}{93.60}$  \\
2.5& 1.0 &    3 & $\underset{(0.630)}{94.14}$&  $\underset{(   1.261)}{93.68}$&   $\underset{(0.177)}{93.86}$&   $\underset{(0.348)}{92.74}$& $\underset{(1.029)}{89.40}$& $\underset{(0.837)}{88.16}$&  $\underset{(0.169)}{93.56}$&  $\underset{(0.344)}{93.40}$  \\
2.5& 1.0 &    4 & $\underset{(0.682)}{94.26}$&  $\underset{(   1.346)}{93.34}$&   $\underset{(0.150)}{93.80}$&   $\underset{(0.299)}{94.06}$& $\underset{(1.118)}{90.24}$& $\underset{(0.948)}{88.96}$&  $\underset{(0.141)}{93.46}$&  $\underset{(0.286)}{93.94}$  \\
\hline
2.5& 0.1 &    1 & $\underset{(0.107)}{95.18}$&  $\underset{(   0.203)}{92.88}$&   $\underset{(0.107)}{94.94}$&   $\underset{(0.213)}{93.92}$& $\underset{(0.176)}{91.18}$& $\underset{(0.157)}{90.90}$&  $\underset{(0.098)}{94.04}$&  $\underset{(0.208)}{94.72}$  \\
2.5& 0.1 &    2 & $\underset{(0.095)}{93.76}$&  $\underset{(   0.208)}{94.42}$&   $\underset{(0.054)}{94.02}$&   $\underset{(0.119)}{94.90}$& $\underset{(0.159)}{89.40}$& $\underset{(0.127)}{88.28}$&  $\underset{(0.035)}{94.50}$&  $\underset{(0.071)}{94.04}$  \\
2.5& 0.1 &    3 & $\underset{(0.616)}{94.42}$&  $\underset{(   1.231)}{93.92}$&   $\underset{(0.108)}{94.64}$&   $\underset{(0.214)}{93.48}$& $\underset{(0.986)}{89.36}$& $\underset{(0.784)}{87.96}$&  $\underset{(0.099)}{94.02}$&  $\underset{(0.208)}{94.80}$  \\
2.5& 0.1 &    4 & $\underset{(0.664)}{94.08}$&  $\underset{(   1.306)}{92.96}$&   $\underset{(0.054)}{94.34}$&   $\underset{(0.117)}{95.08}$& $\underset{(1.081)}{89.90}$& $\underset{(0.901)}{88.70}$&  $\underset{(0.035)}{94.38}$&  $\underset{(0.070)}{94.32}$  \\
 \hline
  \end{tabular}
\end{table}

\begin{table}
 \caption{
Coverage probabilities and average length of  Wald intervals based on
the IPW, SIPW, and ELW estimators
when data were generated from Example \ref{ex2} when $n=500$.
\label{tab-cov-ex2-500} }
\centering
  \begin{tabular}{ | c c|rrr|rrr|}
  \hline
  && \multicolumn{3}{c}{Coverage probability} &  \multicolumn{3}{c|}{Average length}   \\
   $\rho$  &    Model    &    IPW-an &   SIPW-an &   ELW-an  &     IPW-an &   SIPW-an &   ELW-an \\  \hline
 && \multicolumn{6}{c|}{Poisson sampling}   \\\hline
 0.2 &  1&  93.28  &  95.34  & 93.86  &   0.476   &   0.280  &   0.251  \\
 0.2 &  2&  94.26  &  93.18  & 93.52  &   0.507   &   0.333  &   0.275  \\
 0.2 &  3&  93.06  &  95.16  & 93.96  &   1.771   &   0.278  &   0.252  \\
 0.2 &  4&  93.94  &  93.78  & 93.58  &   1.811   &   0.334  &   0.274  \\
 0.8 &  1&  94.72  &  90.90  & 93.52  &   0.431   &   0.343  &   0.232  \\
 0.8 &  2&  96.04  &  91.84  & 95.34  &   0.684   &   0.776  &   0.479  \\
 0.8 &  3&  93.80  &  90.26  & 93.54  &   1.752   &   0.340  &   0.233  \\
 0.8 &  4&  94.66  &  91.78  & 94.92  &   1.928   &   0.778  &   0.479  \\   \hline   && \multicolumn{6}{c|}{Pivotal sampling}   \\  \hline
 0.2 &  1&  96.82  &  94.34  & 92.84  &   0.494   &   0.270  &   0.247  \\
 0.2 &  2&  98.66  &  92.52  & 93.68  &   0.528   &   0.313  &   0.262  \\
 0.2 &  3&  97.86  &  94.84  & 93.40  &   1.789   &   0.267  &   0.246  \\
 0.2 &  4&  98.52  &  92.48  & 93.56  &   1.828   &   0.311  &   0.263  \\
 0.8 &  1&  99.98  &  89.40  & 94.90  &   0.438   &   0.317  &   0.214  \\
 0.8 &  2& 100.00  &  91.46  & 97.48  &   0.698   &   0.729  &   0.436  \\
 0.8 &  3&  98.98  &  88.80  & 94.46  &   1.734   &   0.310  &   0.214  \\
 0.8 &  4&  99.88  &  89.78  & 97.10  &   1.936   &   0.724  &   0.435  \\\hline   && \multicolumn{6}{c|}{PPS sampling}   \\  \hline
0.2  &  1&  92.34  &  94.48  & 91.34  &   0.433   &   0.289  &   0.253  \\
0.2  &  2&  93.54  &  92.96  & 91.80  &   0.451   &   0.333  &   0.262  \\
0.2  &  3&  91.60  &  94.52  & 91.28  &   1.560   &   0.288  &   0.253  \\
0.2  &  4&  91.70  &  93.06  & 92.28  &   1.558   &   0.334  &   0.263  \\
0.8  &  1&  97.52  &  89.46  & 90.64  &   0.311   &   0.333  &   0.188  \\
0.8  &  2& 100.00  &  90.86  & 91.94  &   0.419   &   0.762  &   0.330  \\
0.8  &  3&  92.18  &  89.58  & 91.32  &   1.471   &   0.335  &   0.187  \\
0.8  &  4&  94.34  &  90.56  & 91.04  &   1.468   &   0.775  &   0.330  \\
  \hline
  \end{tabular}
\end{table}

\begin{table}
 \caption{
 Simulated RMSEs of the  IPW, SIPW, ZZZ, and ELW estimators
 where data  were generated from Example \ref{ex2} with $n=500$.
  \label{tab-rmse-ex2-500}
 }
 \tabcolsep 2pt
\centering
  \begin{tabular}{  c c|rrrr|rrrr|rrrr}
  \hline
    &&     IPW &   SIPW & ZZZ      &  ELW &     IPW &   SIPW & ZZZ      &  ELW &     IPW &   SIPW & ZZZ      &  ELW \\  \hline

   $\rho$  &    Model& \multicolumn{4}{c|}{Poisson sampling}  &    \multicolumn{4}{c|}{Pivotal sampling} &  \multicolumn{4}{c}{PPS sampling}   \\\hline
0.2  &   1& 9.04&  4.34&  6.28& {\bf 3.93}&     40.67&  4.38&  5.03& {\bf 3.78}&     7.41 & 4.39&  5.11& {\bf 3.87}  \\
0.2  &   2&11.38&  5.36&  6.95& {\bf 4.16}&     40.64&  5.36&  4.89& {\bf 3.92}&     7.28 & 5.31&  4.94& {\bf 4.03}  \\
0.2  &   3&36.21&  4.59& 23.56& {\bf 3.89}&    179.21&  4.38& 17.48& {\bf 3.78}&    27.46 & 4.39& 18.01& {\bf 3.87}  \\
0.2  &   4&36.55&  5.36& 24.06& {\bf 4.16}&    179.18&  5.36& 17.27& {\bf 3.92}&    27.29 & 5.31& 17.76& {\bf 4.03}  \\ \hline
0.8  &   1& 8.07&  5.66&  5.85& {\bf 3.39}&     24.90&  5.88&  3.07& {\bf 2.91}&     4.53 & 5.74&  3.13& {\bf 3.00}   \\
0.8  &   2&10.79& 12.84&  9.24& {\bf 6.56}&     24.84& 13.29& {\bf 2.79}& 5.03 &     4.27 &13.14& {\bf 2.79}&  5.28   \\
0.8  &   3&33.58&  5.66& 23.09& {\bf 3.39}&    163.52&  5.88& 15.86& {\bf 2.91}&    24.97 & 5.74& 16.36& {\bf 3.00}   \\
0.8  &   4&35.35& 12.84& 25.53& {\bf 6.56}&    163.41& 13.29& 15.03& {\bf 5.03}&    24.32 &13.14& 15.40& {\bf 5.28}   \\   \hline
  \end{tabular}
\end{table}

\begin{table}
 \caption{
 Simulated RMSEs of the IPW, SIPW, ZZZ and ELW estimators
when data were generated from  the {\tt LLvsPSID}  dataset with $n=200$.
  \label{tab-rmse-realdata} }
 \tabcolsep 8pt
 \renewcommand{\arraystretch}{1}
\centering
  \begin{tabular}{  l|cc|cc|cc}
  \hline
  &    IPW  &   ZZZ  &     IPW  &   ZZZ  & SIPW     &  ELW \\
  &   \multicolumn{2}{ c }{$a=0$} & \multicolumn{2}{ c|}{$a=2$} &  &  \\ \hline

 Poisson sampling &   9.35&  8.44& 19.27& 16.33& 8.41& 6.14    \\
 Pivotal sampling &   5.07&  3.91& 12.17&  7.63& 7.15& 4.66   \\
 PPS sampling     &   5.46&  4.13& 13.86&  8.19& 8.70& 5.51    \\    \hline
  \end{tabular}

\end{table}

\end{document}